\documentclass[twocolumn, tighten, times]{aastex63}
\usepackage{amssymb}
\usepackage{amsmath}
\usepackage{bm}
\usepackage{tensor}
\usepackage{braket}
\usepackage{subfigure} 

\newcommand{\SOP}{\affiliation{School of Physics and State Key Laboratory
of Nuclear Physics and Technology, Peking University, Beijing 100871,
China}}
\newcommand{\KIAA}{\affiliation{Kavli Institute for Astronomy and
Astrophysics, Peking University, Beijing 100871, China}}
\newcommand{\DoA}{\affiliation{Department of Astronomy, School of Physics,
Peking University, Beijing 100871, China}}

\newcommand{\NAOC}{\affiliation{National Astronomical Observatories,
Chinese Academy of Sciences, Beijing 100012, China}}
\newcommand{\CICQM}{\affiliation{Collaborative Innovation Center of Quantum
Matter, Beijing, China}}
\newcommand{\CHEP}{\affiliation{Center for High Energy Physics, Peking
University, Beijing 100871, China}}

\received{June 1, 2021}
\revised{June 1, 2021}
\accepted{June 1, 2021}
\submitjournal{ApJ}
\shorttitle{LISA-SKA Pulsars}
\shortauthors{X. Miao {\it et al.}}
\begin{document}

\title{Stringent Tests of Gravity with Highly Relativistic Binary Pulsars
in the Era of LISA and SKA}
\correspondingauthor{Lijing Shao}
\email{lshao@pku.edu.cn}
\author[0000-0003-1185-8937]{Xueli Miao}\SOP
\author{Heng Xu}\DoA\KIAA
\author[0000-0002-1334-8853]{Lijing Shao}\KIAA\NAOC
\author[0000-0001-7649-6792]{Chang Liu}\DoA\KIAA
\author[0000-0002-5660-1070]{Bo-Qiang Ma}\SOP\CICQM\CHEP

%---------------------------------------------------------------------
\begin{abstract}

At present, 19 double neutron star (DNS) systems are detected by radio timing
and 2 merging DNS systems are detected by kilo-hertz gravitational waves.
Because of selection effects, none of them has an orbital period $P_b$ in the
range of a few tens of minutes. In this paper we consider a multimessenger
strategy proposed by \citet{Kyutoku:2018aai}, jointly using the Laser
Interferometer Space Antenna (LISA) and the Square Kilometre Array (SKA) to
detect and study Galactic pulsar-neutron star (PSR-NS) systems with $P_b \sim$
10--100\,min. We assume that we will detect PSR-NS systems by this strategy.  We
use standard pulsar timing software to simulate times of arrival of pulse
signals from these binary pulsars. We obtain the precision of timing parameters
of short-orbital-period PSR-NS systems whose orbital period $P_b \in
(8,120)\,$min. We use the simulated uncertainty of the orbital decay,
$\dot{P}_{b}$, to predict future tests for a variety of alternative theories of
gravity.  We show quantitatively that {highly} relativistic PSR-NS systems
will significantly improve the constraint on parameters of specific gravity
theories in the strong field regime. We also investigate the orbital periastron
advance caused by the Lense-Thirring effect in a PSR-NS system with $P_b =
8\,$min, and show that the Lense-Thirring effect will be detectable to a good
precision.

\end{abstract}
%---------------------------------------------------------------------
%---------------------------------------------------------------------

% \tableofcontents

%---------------------------------------------------------------------
\section{Introduction} 
\label{sec:intro}
%---------------------------------------------------------------------

Currently, general relativity (GR) is the best tested theory of gravity and has
passed all the tests with flying colors~\citep{Will_2014}.  But there are
tentative alternative theories that could possibly explain the ``missing mass''
problem in galaxies and the accelerating expansion of our Universe without
introducing dark matter or dark energy \citep{Jain_2010,Clifton_2012}. In
addition, there are theoretical arguments that GR may not be the exactly correct
gravity theory in the ultraviolet end, and it is necessary to test GR and
alternative theories more precisely. There are many gravity tests in the Solar
System and these tests provide very tight restriction on the parameter space of
theories beyond GR.  Great achievements were made in bounding the parametrized
post-Newtonian (PPN) parameters \citep{Will:2018bme}.  Because different
theories of gravity correspond to different values of PPN parameters in the weak
field, bounding PPN parameters can help us test theories of gravity in a
systematic way.  The Solar System experiments probe gravitation in the weak
field regime.  However, for some gravity theories, such as the scalar-tensor
gravity of \citet{Damour:1993hw}, gravity
can be reduced to GR in the weak field, while it is very different from GR in
the strong field \citep{Damour_1992, Damour_1996, Shao:2016ezh}.  So testing
theories of gravity in the strong field can help us understand gravity theory
more comprehensively.

Radio pulsars, rotating neutron stars (NSs) with radio emission along their
magnetic poles, are ideal astrophysical laboratories for strong-field gravity
tests, because of the strong self-gravity of NSs. Binary pulsar systems, in
which one or more bodies are NSs, provide us an opportunity to test gravity in
the strong field. We can measure the parameters of binary pulsar systems with
very high precision, which benefits from an extremely accurate measurement
technology, the so-called pulsar timing \citep{2004hpa..book.....L,Wex:2014nva}.
Pulsar timing precisely measures the times of arrival (TOAs) of radio signals at
the telescope on the Earth and fits an appropriate timing model to these TOAs to
get a phase-connected solution. In 1974, \citet{Hulse:1974eb} discovered the
first binary pulsar system, PSR~B1913+16, which is a double NS (DNS) system.
This system proved the existence of gravitational wave (GW) radiation for the
first time \citep{Taylor:1979zz}, using the fact that the observed orbital
period decay rate, $\dot{P}_{b}^{\rm obs}$, is consistent with the predicted
value from GR, $\dot{P}_{b}^{\rm GR}$. With more and more binary systems
discovered, we can use them to test a variety of theories of gravity~\citep[see
e.g.][]{Manchester:2015mda, Kramer:2016kwa, Shao:2012eg, Shao:2013eka,
Shao:2013wga, Shao:2014oha, Shao:2014bfa, Miao:2020wph, Xu:2020vbs, Xu:2020zxs}.
For example, $\dot{P}_{b}$ of binary pulsar systems can be used to restrict the
time variation of the gravitational constant $G$ \citep{Wex:2014nva,Zhu_2018};
it can also be used to bound the mass of graviton in massive graviton theories
\citep{Finn_2002, Miao:2019nhf, Shao:2020exw}.

At present, more than 300 binary pulsar systems have been
discovered\footnote{https://www.atnf.csiro.au/research/pulsar/psrcat/}
\citep{ATNFdata}, including pulsar-NS (PSR-NS) systems, pulsar-white dwarf
(PSR-WD) systems, pulsar-main sequence star (PSR-MS) systems and so on.  Among
them, PSR-NS systems are ideal to test gravity theory, because most pulsars in
these systems are millisecond pulsars (MSPs) which come from the ``reversal
mechanism'' \citep[namely, the so-called ``recycled
pulsars'';][]{Tauris_2017}.  MSPs are generally more precise timers than
normal pulsars, because the timing precision of normal pulsars are usually
dominated by red noise, spin irregularities, glitched, etc. \citep{Lyne_2010}.
Therefore MSPs can provide more precise gravity tests.

Up to now, 19 DNS systems have been discovered from radio
telescopes\footnote{Several companion stars of these systems are not fully
identified as NSs.} \citep{ATNFdata}.  One of them is a double pulsar system,
PSR~J0737$-$3039A/B \citep{Kramer:2006nb}, which provides a wealth of
relativistic tests.  The DNS system with the shortest orbital period is
PSR~J1946+2052 with ${P}_{b}=0.078\,{\rm day}$, and this system has the largest
periastron advance, $\dot{\omega}=25.6\pm0.3\,{\rm deg\,yr^{-1}}$
\citep{Stovall:2018ouw}. {Binary pulsar systems with shorter orbital
periods can provide more relativistic gravity tests because of higher relative
velocity $v$.  They also can provide smaller fractional uncertainties for some
binary orbital parameters. For example, the expected dependence of fractional
uncertainty of $\dot{P}_{b}$ is proportional to $P_{b}^{3}$
\citep{Damour:1991rd,2004hpa..book.....L}.  So in most cases shorter orbital
period PSR-NS systems can help us test gravity better. But we} have not detected
DNS systems with orbital periods of several minutes by radio observation yet,
because the fast-changing Doppler shift caused by orbital acceleration of the
pulsar can smear the pulsar signal in the Fourier domain. Therefore, the
selection effect makes the detection of pulsars in very tight binary systems
difficult \citep{Tauris_2017}.

In addition to radio observations, GW observations from LIGO/Virgo have also
detected two DNS systems, GW170817 in the second observing run \citep[O2;
][]{TheLIGOScientific:2017qsa} and GW190425 in the third observing
run~\citep[O3; ][]{Abbott:2020uma}.\footnote{The possibility that one or both
components of {GW190425} are black holes cannot be fully ruled out. 
{For GW170817, a BH–NS merger is also possible, but it is more likely to be
a DNS merger \citep{Coughlin_2019}.}} Because of the high frequency range of
LIGO and Virgo, we can only detect GWs from DNS systems near their phase of
merger.  Therefore, the orbital period of the discovered DNSs has a gap between
1.88$\,$hr (PSR~J1946+2052) and $\sim 1\,$ms at the phase of merger.  If we can
detect a DNS system with an orbital period in this gap, it can help us better
understand the population and formation mechanism of DNS systems.  On the other
hand, these systems with orbital periods of several minutes are highly
relativistic, so they can provide complementary gravity tests to what have been
done with current pulsar and GW observations.

In 2018, \citet{Kyutoku:2018aai} proposed a multimessenger strategy, combining
the Laser Interferometer Space Antenna (LISA) and the Square Kilometre Array
(SKA), to detect radio pulses from Galactic DNSs with a period shorter than 10
minutes.  LISA is a future space-based GW detector, and it is sensitive at mHz
bands \citep{Audley:2017drz}, which makes DNS systems with ${P}_{b}\sim\,$min a
target for detection.  SKA (Phase 2) under construction will greatly improve
detection sensitivity of radio observation in the
future.\footnote{https://astronomers.skatelescope.org/ska/} According to the
schedule, LISA and SKA will perform their missions simultaneously for a duration
of time.  \citet{Kyutoku:2018aai} considered that we can utilize LISA to
discover DNS systems, and provide their  sky location and binary parameters with
a high accuracy.  Then one can use SKA to probe whether or not a visible pulsar
exists in the DNS system.  The sky location information can help SKA locate the
DNS, and the information of binary parameters can help recover pulse signals
modulated by orbital motion.  With the help of LISA, SKA can detect PSR-NS
systems with ${P}_{b}\sim\,$min more easily with affordable computational cost.
So to some extent, the multimessenger strategy can help us reduce the difficulty
in searches of short-orbital-period PSR-NS systems.

As a future high-sensitivity radio instrument, SKA can improve the measurement
precision of TOAs of pulse to an unprecedented level, which can help improve
the measured accuracy of binary pulsar parameters.  The high precision binary
parameters can further help us improve gravity tests.  The theoretically
projected relation between $P_{b}$ and the fractional error of $\dot{P}_{b}$
in measurement is $\delta(\dot{P}_{b})\propto P_{b}^{3}$
\citep{2004hpa..book.....L}.  So for the short-orbital-period PSR-NS systems,
$\delta(\dot{P}_{b})$ can be well measured.  Then we can use the high
precision $\dot{P}_{b}$ to provide tighter bounds on the parameter space of
gravity theories beyond GR.

In general, short-orbital-period binary pulsars are short lived.  Recent
literature have estimated the number of DNS systems that can be detected by
LISA.  Based on different Galaxy DNS merger rates, \citet{Andrews:2019plw}
considered that LISA will detect on average 240 (330) DNS systems in the Galaxy
for a 4-yr (8-yr) mission with signal to noise ratio (SNR) above 7.
\citet{Lau:2019wzw} estimated that around 35 DNSs will accumulate SNR$\,>8$ over
a 4-yr LISA mission.  Although the estimated number for DNSs to be detected by
LISA are different in the above two papers, they all confirm that a certain
number of DNS systems are likely to be detected by LISA.  Therefore, we expect
that the multimessenger strategy can detect short-orbital-period PSR-NS systems
in the future.

In this work, we assume that in the future some short-orbital-period PSR-NS
systems are detected by the multimessenger strategy.  Based on this assumption,
we simulate the fractional error of parameters of these short-orbital-period
PSR-NS systems using the standard pulsar timing software
TEMPO2,\footnote{https://bitbucket.org/psrsoft/tempo2} and use the simulated
precision of the parameter $\dot{P}_{b}$ to constrain specific theories of
gravity, including varying-$G$ theory, massive gravity and scalar tensor
theories.  Results of our simulation show significant improvement in the
constraints of specific parameters of gravity theories in the strong field.  We
also provide a simple estimate of  orbital periastron advance rate, $\dot{\omega}_{\rm
LT}$, due to the Lense-Thirring effect.  Our result suggests that we are able to
measure the value of $\dot{\omega}_{\rm LT}$ in the short-orbital-period PSR-NS
systems, which can help us restrict the equations of state (EOSs) of NSs.  We
need to underline that we have ignored higher order contributions in our
simulation, thus our results are indicative, but providing a reasonable estimate
of the magnitude of the precision of parameters. 

The paper is organized as follows.  In the next section, we overview the
multimessenger strategy and list the estimated number of DNSs which could be
detected by LISA from several work.  In Sec.\,\ref{sec:sim}, we simulate the
parameters of PSR-NS systems to be detected by the joint observation of LISA and
SKA.  We give the simulated fractional uncertainty of parameters.  In
Sec.\,\ref{sec:test:GR}, we use our simulated fractional uncertainty of the
parameter $\dot{P}_{b}$ to constrain parameter spaces of some specific theories
of gravity.  In Sec.\,\ref{sec:lense}, we provide an estimate on the order of
magnitude of orbital periastron advance rate, $\dot{\omega}_{\rm LT}$.  Finally
we give a discussion in Sec.\,\ref{sec:disc}.

%---------------------------------------------------------------------
\section{Observing DNSs with LISA and SKA}
\label{sec:lisa:ska}
%---------------------------------------------------------------------

At present, the detected DNS system of the shortest orbital period is PSR
J1946+2052 with $P_{b}=0.078\,{\rm day}$ \citep{Stovall:2018ouw}.  The merging
timescale is provided by \citep{PhysRev.136.B1224,Liu:2014uka},
%------------------
\begin{equation}
     T_{\rm merge}=\frac{5}{256}\left(\frac{P_{b}}{2\pi}\right)^{8/3}T_{\odot}^{-5/3}\frac{(m_{p}+m_{c})^{1/3}}{m_{p}m_{c}}G(e)\,,
    \label{eq:Tmerge}
\end{equation}
%------------------
where
%------------------
\begin{equation}
    G(e)=1- 3.6481e^{2}+5.1237e^{4}- 3.5427e^{6}+1.3124e^{8}+\cdots\,.
    \label{eq:Ge}
\end{equation}
%------------------
In Eq.~(\ref{eq:Tmerge}), $T_{\odot}\equiv GM_{\odot}/c^{3}\approx4.9254909\,{\rm \mu
s}$ where $M_{\odot}$ is the Solar mass, $m_{p}$ and $m_{c}$ are the masses of
the pulsar and the companion in Solar units respectively, and $e$ is the
eccentricity of the binary system.  By calculating Eq.\,(\ref{eq:Tmerge}), we
get the merging timescale of PSR J1946+2052, $T_{\rm merge}=46\,{\rm Myr}$,
which means that this system will take $46\,{\rm Myr}$ to merge.  The GW events
from DNS merging suggest that the Galactic merger rate of DNSs is about one in
$10^{4}\,{\rm yr}$ \citep{TheLIGOScientific:2017qsa,Abbott:2020uma}.  So there
must be undetected DNS systems with $P_{b}<0.078\,{\rm day}$ in the Galaxy.  The
reason why we have not observed DNS systems with orbital period less than one
hour is that we are constrained by our present searching methods.  Radio
observations contributed mostly to DNS detection, as 19 DNS systems have been
detected by radio telescopes.  Orbital motion of the short-orbital-period
binary systems can lead to a severe Doppler smearing which makes it difficult
for radio observation to recover the pulse signals of such systems.  In GW
observation, LIGO/Virgo detected two DNS systems, GW170817 in O2 and GW190425
in O3 \citep{TheLIGOScientific:2017qsa,Abbott:2020uma}.  However, LIGO/Virgo
are only sensitive to the merger phase.  Therefore, constrained by the
existing technology and observation, currently it is difficult to detect DNS
systems with $P_{b}\sim{\rm min}$. 

LISA, a future space-based GW detector, which is expected to be operational in
2030s, is sensitive at mHz bands, so the DNS systems with
${P}_{b}\sim{\rm min}$ are potential targets for LISA.  Recently, several works
have provided estimates of the number of DNS systems that could be detected by
LISA.  \citet{Andrews:2019plw} used the Galaxy DNS merger rate of $210\,{\rm
Myr^{-1}}$,  which is derived from \citet{TheLIGOScientific:2017qsa,Abbott_2019prx}, to estimate the number of
DNSs that LISA could detect.  They showed that LISA will detect on average 240
(330) DNS systems in the Galaxy with ${\rm SNR>7}$ for a 4-yr (8-yr) mission.
They also used a conservative DNS merger
rate, $42^{+30}_{-14}\,{\rm Myr^{-1}}$ \citep{Pol_2019}, and estimated that LISA
will detect on average 46 (65) DNS systems in the Galaxy for a 4-yr (8-yr)
mission with ${\rm SNR>7}$.  The smaller merger rate 
{could suffer from a selection effect \citep{Tauris_2017}.}
At the same time, \citet{Lau:2019wzw} {provided an estimation} that around 35 DNSs will accumulate SNR above 8 over a 4-yr LISA
mission.  They gave a smaller detectable number
of DNSs than that in \citet{Andrews:2019plw}.  It is because that
\citet{Lau:2019wzw} used a smaller DNS merger rate, $33 \,{\rm Myr^{-1}}$. Overall, no
matter we use a conservative or an optimistic DNS merger rate, LISA has a high
probability of detecting the DNS systems with $P_{b}\sim{\rm min}$, and there
may be PSR-NS systems within them.

\citet{Kyutoku:2018aai} provided a multimessenger strategy, {using LISA and SKA,} to detect PSR-NS
systems with $P_{b} \sim 10\,{\rm min}$ in the Galaxy.  {LISA} can detect DNSs with $P_{b}\sim\,{\rm min}$ and
provide information on the location and orbital parameters of DNSs. These
information can help SKA locate DNS systems and recover the radio pulse signal
which could exist in the system \citep{Kyutoku:2018aai}. So the joint
observation strategy can significantly reduce the number of trials used to
search for DNS systems with radio telescopes alone.
\citet{Kyutoku:2018aai} estimated the 1-$\sigma$
statistical errors of parameters {when SNR $\rho=200$ for a 2-yr observation mission,} {and the uncertainties of parameters can be found in their Eqs.\,(9-11).}
With the estimated errors of parameters, \citet{Kyutoku:2018aai} showed {that LISA can locate DNS within
an error range $\Delta\Omega\approx0.036\,{\rm deg^{2}}$ [see their Eq. (11)] and then SKA only needs to take 240 pointings to locate the DNS system
precisely by a
tied-array beam.}
So the multimessenger strategy can greatly reduce the number of
points that an all-sky survey needs.  After locating the DNS system, {we can use the orbital frequency and binary parameters provided by LISA to recover possible pulse signal of the pulsar, which also can greatly reduce the number of required trials when using a coherent integration.\footnote{The specific comparisons are shown in Sec.\,3.2 of \citet{Kyutoku:2018aai}.}}
We should notice that all-sky survey is not
confined to detect PSR-NS systems with $P_{b}\sim10\,{\rm min}$. It can be
used for other radio source searching as well.  The multimessenger observation
can specifically {and efficiently} help us detect PSR-NS systems with $P_{b}\sim10\,{\rm min}$.

Based on \citet{Kyutoku:2018aai}, \citet{Thrane:2019lwv} offered an extended
analysis for the joint detection of LISA and SKA.  They considered a 4-yr
observation mission of LISA where the SNR can be
improved to $\rho=360$, {so} the fractional errors of parameters can be {further} reduced.
With the improved precision of parameters, we can use the joint detection to
detect short-orbital-period PSR-NS systems more efficiently.

In conclusion, with a reasonable merger rate for Galactic DNSs, the joint
detection of SKA and LISA can provide an efficient and more targeted way for
detecting PSR-NS systems with very short orbital period.

%---------------------------------------------------------------------
\section{Pulsar-timing simulations}
\label{sec:sim}
%---------------------------------------------------------------------

The multimessenger strategy can offer a specific approach to seek for the
short-orbital-period PSR-NS systems.  
{These systems can provide a more relativistic situation to test gravity.}
In our paper, we assume that we will
have detected PSR-NS systems by this way in the future, and then we simulate
the precision of orbital parameters of the short-orbital-period PSR-NS systems
for the SKA.

SKA as the next generation radio telescope will enable the precision of TOAs to
reach an unprecedented level.  \citet{Liu:2014uka} provided approximated
measurement precision on TOAs at $1.4\,{\rm GHz}$ {with 10-min integration}, which is listed in
Table\,{1 of their work}.  These results are based on the presumed instrumental
sensitivities.  In practice, the precision of TOAs depends not only on
instrumental sensitivities, but also on interstellar medium evolution, intrinsic
pulse shape changes, etc. \citep{Liu:2011cka}.  Based on various effects,
\citet{Liu:2011cka} predicted that a TOA precision of a normal-brightness MSP
can achieve $80$--230$\,{\rm ns}$ at $1.4\,{\rm GHz}$ with 10-min integration by
the SKA.

{The timing precision of young pulsars is normally dominated by red noise,
spin irregularities and glitches, etc., so the timing precision could not be
significantly improved by increasing the system sensitivity of telescopes.
Hence we assume that the pulsar in PSR-NS system is recycled, which has stable
precision of TOAs in general, and we use a spin period $P=20\,{\rm ms}$.  For
simplification, we only consider white noise.} In our work, considering the
possible contributions of various effects, we use two different measurement
precisions of TOAs to perform the simulations, $\sigma_{\rm TOA}=100\,{\rm ns}$
and $\sigma_{\rm TOA}=1\,\mu {\rm s}$.  LISA is sensitive to DNS systems with
${P}_{b}\sim10\,{\rm min}$, so the target of multimessenger observation strategy
is mainly PSR-NS systems with ${P}_{b}\sim10\,{\rm min}$.  Meanwhile, radio
observation can also provide us opportunity to detect systems with a period
around {100} min by ``acceleration search'' {\citep{1991ApJohnston}}
and ``jerk search'' {\citep{Bagchi_2013,Andersen_2018}}.  So we simulate
PSR-NS systems with $P_{b}\in [8,\,120]\,{\rm min}$.  For the masses of the NSs,
we choose two sets of parameters, $m_{p}=1.3\,M_{\odot}$, $m_{c}=1.7\,M_{\odot}$
and $m_{p}=1.35\,M_{\odot}$, $m_{c}=1.44\,M_{\odot}$.  \citet{Andrews:2019plw}
used the equations from \citet{PhysRev.136.B1224} to show how orbits of the 20
known PSR-NSs\footnote{Their 20 PSR-NSs contain PSR J0514$-$4002A
\citep{Ridolfi_2019} whose companion in ATNF is not explicitly identified as a
NS.} will evolve over the next 10 Gyr as gravitational radiation causes them to
circularize and decay.  Figure\,1 in \citet{Andrews:2019plw} shows that the
orbital eccentricity goes smaller than 0.1 when the orbital period $P_{b}$ of
the 20 PSR-NS systems evolve to 10\,min.  So in our simulation, we set
eccentricity $e=0.1$ {for the simulation of all
systems}.{\footnote{The simulated results also depend on
the value of eccentricity. If we select $e=0.5$ to simulate all systems, the
results will be further improved by 3 to 10 times for different parameters.
Especially for $\dot{\omega}$, its fractional error becomes smaller nearly by an
order of magnitude. Therefore, our results are conservative in this respect.}}

We use the FAKE plug-in in TEMPO2 \citep{Hobbs:2012na} to simulate pulsar timing
data. We need to provide a parameter file which contains parameters of the
pulsar and binary orbit as input.  {In this work, we are interested in how
the short orbit PSR-NS systems combined with the high-precision measurement of
SKA can improve the measurement precision of orbital parameters.  Hence, we only
consider simulating the time delay of TOAs related to orbital motion.  These
time delay terms of orbital motion are the R$\rm{\ddot{o}}$mer delay caused by
the orbital motion of the pulsar, the Einstein delay caused by the gravitational
redshift and the second order Doppler effect, and the Shapiro delay, which
is due to the gravitational field of the companion star.\footnote{{The specific forms of those delay terms can be found in Eqs.\,(2.2b--2.2d) in \citet{Damour:1991rd}.}}} For timing model
{of orbital motion}, different models have different orbital parameters to
fit.  \citet{Damour1986} introduced a phenomenological parametrization, which is
called the parametrized post-Keplerian (PPK) formalism, to describe the orbital
motion of pulsars.  The post-Keplerian (PK) parameters are solutions of the
first order post-Newtonian (PN) equations of orbital motion and it is generic
for fully conservative gravity theories \citep{Damour1985,Damour1986}.  In our
simulation, we choose the Damour-Deruelle (DD) timing model.

In the DD timing model, we are mainly able to measure 5 PK
parameters---$\dot{\omega},\,\gamma,\,r,\,s,\,\dot{P}_{b}$---in binary pulsar
systems.  In GR, they can be expressed as functions of two component masses 
\citep{2004hpa..book.....L},
\begin{align}
  \dot{\omega} &=3 T_{\odot}^{2 / 3}\left(\frac{P_{b}}{2 \pi}\right)^{-5 / 3} \frac{1}{1-e^{2}}\left(m_{p}+m_{c}\right)^{2 / 3}\,, \label{eq:omegadot}\\
  \gamma &=T_{\odot}^{2 / 3}\left(\frac{P_{b}}{2 \pi}\right)^{1 / 3} e \frac{m_{c}\left(m_{p}+2 m_{c}\right)}{\left(m_{p}+m_{c}\right)^{4 / 3}}\,, \label{eq:gamma}\\
  r &=T_{\odot} m_{c}\,, \label{eq:r}\\
  s &=\sin i=T_{\odot}^{-1 / 3}\left(\frac{P_{b}}{2 \pi}\right)^{-2 / 3} x \frac{\left(m_{p}+m_{c}\right)^{2 / 3}}{m_{c}}\,, \label{eq:s}\\
  \dot{P}_{b} &=-\frac{192 \pi}{5} T_{\odot}^{5 / 3}\left(\frac{P_{b}}{2 \pi}\right)^{-5 / 3} f(e) \frac{m_{p} m_{c}}{\left(m_{p}+m_{c}\right)^{1 / 3}}\,, \label{eq:pbdot}
\end{align}
with
\begin{equation}
    f(e) \equiv \frac{1+(73/24)e^{2}+(37/96)e^{4}}{(1-e^{2})^{7/2}}\,. \label{eq:fe}
\end{equation}
In above equations, $x\equiv(a/c)\sin{i}$ is the projected semi-major axis of
the pulsar orbit and $i$ is the orbital inclination; $\dot{\omega}$ is the
relativistic precession of periastron of pulsar; 
{$\gamma$ is a PK parameter which is related to the Einstein delay; $r$ and $s$ are PK parameters which are related to the Shapiro delay;}
$\dot{P}_{b}$ is the orbital
period decay rate caused by GW quadrupole radiation in GR.  {For the 5 PK
parameters, except that parameters $r$ and $s$ do not depend on $P_{b}$, the
expected fractional uncertainty of $\dot{P}_{b}$ is proportional to $P_{b}^{3}$,
the expected fractional uncertainty of $\dot{\omega}$ is proportional to
$P_{b}$, and the expected fractional uncertainty of $\gamma$ is proportional to
$P_{b}^{4/3}$; these dependent relationships are listed in Table II in
\citet{Damour:1991rd}.  So PSR-NS systems with $P_{b}\sim 10\, {\rm min}$ can
provide better measurement precisions for PK parameters $\dot{P}_{b}$,
$\dot{\omega}$ and $\gamma$, which can help us improve the ability of gravity
test.} In principle, there are extra relativistic PK parameters, e.g.,
$\delta_\theta$. We have omitted them in this work.

\begin{deluxetable}{p{3cm}p{2.3cm}p{2.3cm}}
  \tablecaption{Parameters of two example PSR-NS systems. Notice that $m_p$
  and $m_c$ are in unit of the Solar mass, as defined in text.
  \label{tab:PSR-NSsparameters}}
  \tablehead{~ &PSR-NS I & PSR-NS II}
  \startdata
    $m_{p}$ & 1.3 & 1.35\\
    $m_{c}$ & 1.7 & 1.44\\
    Eccentricity, $e$ & 0.1 & 0.1\\
    Orbital inclination, $i$ & $60^{\circ}$ & $60^{\circ}$\\
    $\sigma_{\rm TOA}$ & $100\,{\rm ns}$ \& $1\,\mu {\rm s}$ & $100\,{\rm ns}$ \& $1\,\mu {\rm s}$\\
 \enddata
\end{deluxetable}

As mentioned previously, we should provide the orbital parameters of PSR-NS
systems for the parameter file. We use the parameters of example PSR-NSs listed
in Table\,\ref{tab:PSR-NSsparameters} and
Eqs.\,(\ref{eq:omegadot}--\ref{eq:pbdot}) to provide the values of PK
parameters. In fact, the PK parameters should be generic, however current
measurements of  PK parameters in binary pulsars show no significant deviation
from the predicted values in GR.  In general, we use the error of the
observations of PK parameters as a deviation from GR to limit parameter spaces
of alternative theories of gravity.  Here we assume that GR is correct and
predict the bounds on other theories of gravity constrained by these PSR-NS
systems.  

%--------------------------------------------------------
\begin{figure*}
    \includegraphics[width=14.5cm]{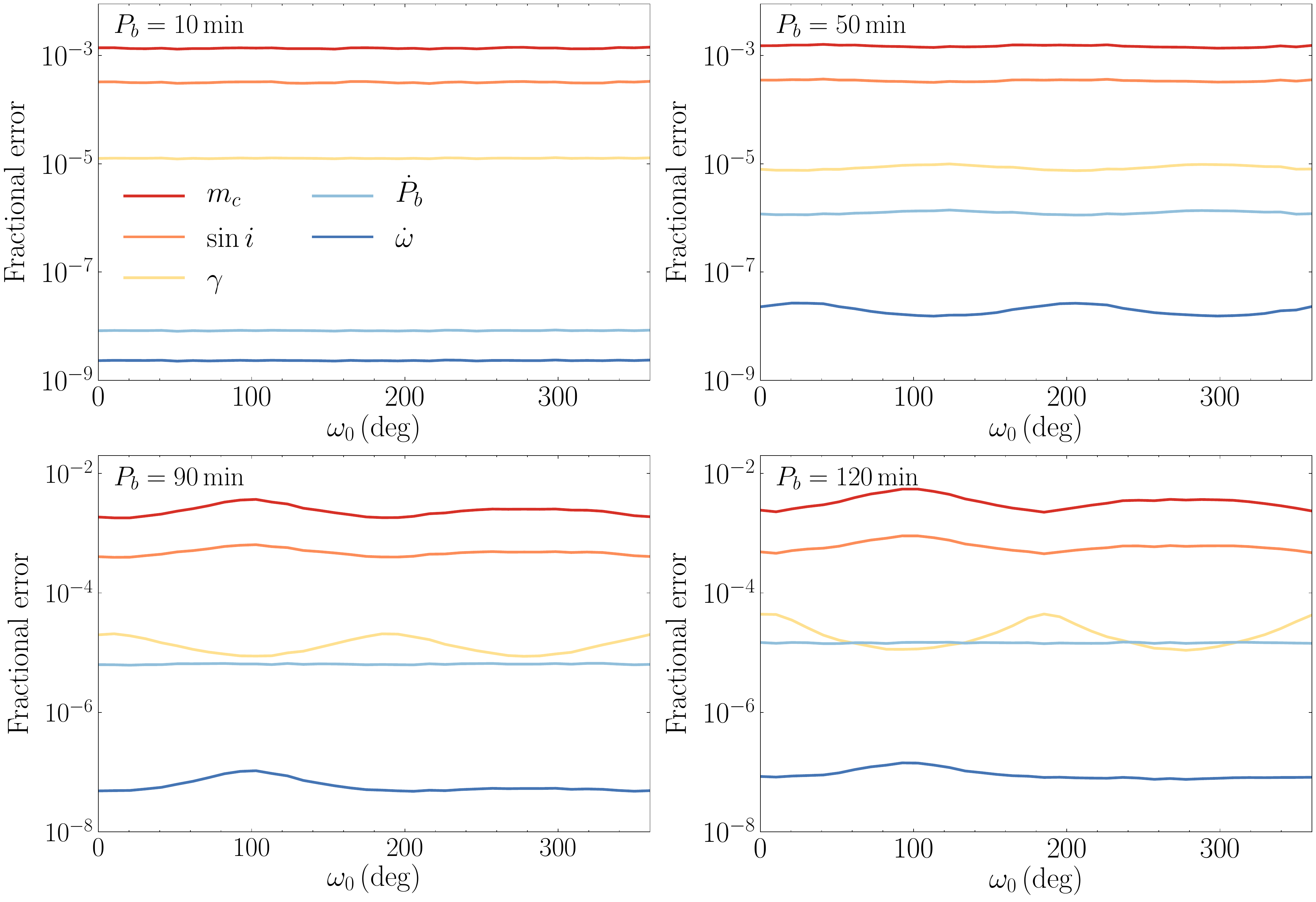}
    \centering
    \caption{The fractional uncertainties of 5 PK parameters vary with the
    value of $\omega_{0}$ for different $P_{b}$,  given in the
    upper left corner of each panel.  The masses of the simulated PSR-NS
    system are $m_{p}=1.3\,M_{\odot},~m_{c}=1.7\,M_{\odot}$, and we have
    assumed $\sigma_{\rm TOA}=100\,{\rm ns}$. \label{fig:omegachange}}
\end{figure*}
%--------------------------------------------------------

We assume a 4-yr observation and ten TOAs per week for the SKA.  {Ten TOAs
per week means that we need to observe a PSR-NS system for 100 min per week,
when each TOA is obtained with a 10-min integration time.\footnote{For PSR-NS
systems with $P_{b}\sim{\rm min}$, we can not use TOAs with integration time
$t_{\rm inte}\sim{\rm min}$, because it can erase the information of orbital
motion, so in practice, the simple treatment here should be improved, and we
need to reduce the integration time of TOAs and increase the number of TOAs in
order to guarantee the equivalent accuracy of timing.} With parameter file and
fitting conditions determined, TEMPO2 can generate simulated TOAs and give a
phase-connected solution using the FAKE plug-in by weighted least-squares
algorithm \citep[see details, see][]{Hobbs_2006}.} The simulation can provide
values and measurement uncertainties of parameters, and we can utilize the
simulated results to predict how the short-period PSR-NS systems can limit
specific theories of gravity.

In our simulation, we need to give a value of the longitude of periastron
$\omega_{0}$ at the epoch of periastron passage $T_{0}$ in parameter file. 
Figure\,3 of \citet{Damour:1991rd} shows that the fractional uncertainties of
5 PK parameters vary with the value of $\omega_{0}$.  The reason is that there
are some degeneracy between different time delay terms, e.g. R\"omer delay and
Einstein delay, and different values of $\omega_{0}$ can lead to different
degrees of degeneracy.  Therefore we simulate the fractional uncertainties of
5 PK parameters for different $\omega_{0}$'s.  We
exhibit results for the ``PSR-NS I'' system with
$\delta_{\rm TOA}=100\,{\rm ns}$ (see Table~\ref{tab:PSR-NSsparameters}) for different $P_{b}$'s in Fig.\,\ref{fig:omegachange}.  As
$P_{b}$ increases, different values of ${\omega}_{0}$ can result in a
significant difference in the fractional errors of 4 PK parameters, except
$\dot{P}_{b}$.  So we take the effect of $\omega_{0}$ into account in the
simulation and calculate the fractional error with
${\omega}_{0}\in\left[0^{\circ},\,360^{\circ}\right)$.

%--------------------------------------------------------
\begin{figure*}
    \includegraphics[width=14.5cm]{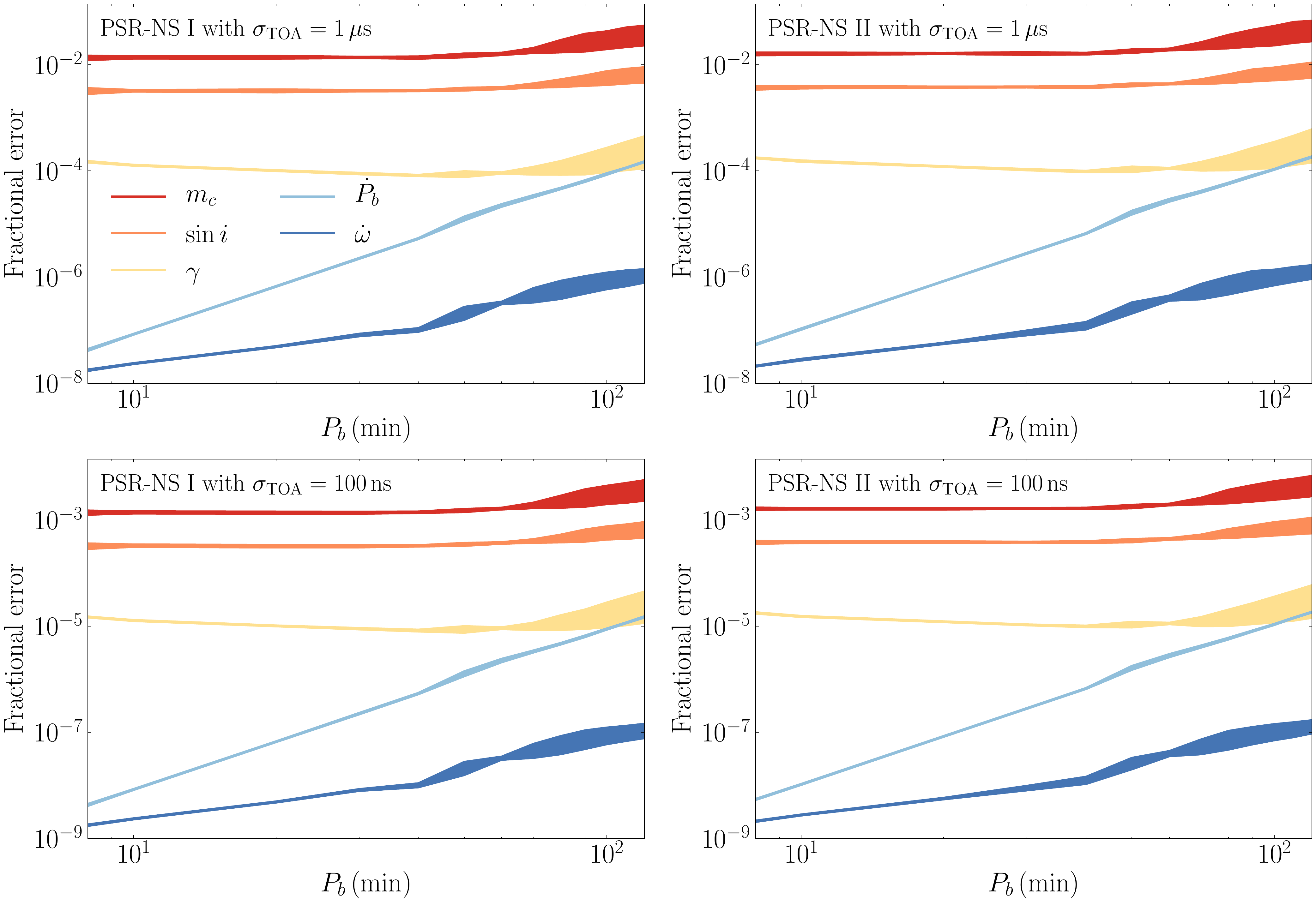}
    \centering
    \caption{The fractional uncertainties of 5 PK parameters as functions of
    $P_{b}$.  Upper and lower panels are results with $\sigma_{\rm
    TOA}=1\,{\rm \mu s}$ and $\sigma_{\rm TOA}=100\,{\rm ns}$ respectively.
    Left and right panels are respectively for PSR-NS I and PSR-NS II (see
    Table~\ref{tab:PSR-NSsparameters}).  The band in the figure is due to
    different values of $\omega_{0}$.  \label{fig:simulateresulte}}
\end{figure*}
%--------------------------------------------------------

{Figure\,\ref{fig:simulateresulte} shows the fractional uncertainties of 5 PK parameters as functions of $P_{b}$ for a 4-yr observation.}
The band of the expected fractional uncertainty is caused by simulating with
different $\omega_{0}$'s.  In the upper panels of
Fig.\,\ref{fig:simulateresulte}, we present the expected measurement precisions
of the 5 PK parameters with $\sigma_{\rm TOA}=1\,{\rm \mu s}$.  In the lower
panels of Fig.\,\ref{fig:simulateresulte}, we present the results with
$\sigma_{\rm TOA}=100 \,{\rm ns}$.  
{As shown in Fig.\,\ref{fig:simulateresulte},
the simulated fractional uncertainties of PK parameters with $\sigma_{\rm
TOA}=100 \,{\rm ns}$ are an order of magnitude smaller than those with
$\sigma_{\rm TOA}=1\,{\rm \mu s}$.  SKA can provide a high precision measurement
for TOAs of pulsars. The expected sensitivity could be even smaller than 100 ns
\citep{Liu:2014uka}, so the precision of PK parameters might be even higher for real
detection in the future.  At the same TOA precision, the simulated results of
the PSR-NS I system are slightly better than that of the PSR-NS II system, which
could be caused by a more significant mass difference and thus some reduced
degeneracy among parameters.}

{Based on a 4-yr observation, we} notice that the simulated fractional
uncertainty of $\dot{P}_{b}$ as a function of $P_{b}$ is largely consistent
with what theoretically predicts, namely $\propto P_{b}^{3}$
\citep{Damour:1991rd,2004hpa..book.....L}.\footnote{ {There is still a
slight deviation from the theoretical prediction in interval
$P_{b}\in[40,\,60]\,{\rm min}$, which could be due to slight parameter
dependence and the validity of the simplified theoretical prediction.} }  { So the
predicted fractional uncertainty of $\dot{P}_{b}$ can offer reasonable
predictions of gravity tests. The simulated fractional uncertainties of PK
parameters $r$ and $s$ are mainly consistent with the theoretical prediction,
which are not dependent on $P_{b}$, except that there is a slight rise in
$P_{b}\in[40,\,60]\,{\rm min}$.  We notice that the slope of PK parameter
$\dot{\omega}$ changes significantly in $P_{b}\in[30,\,60]\,{\rm min}$ for a
4-yr observation, which could be caused by the degeneracy between different
time delay terms.  For parameter $\gamma$ in a 4-yr observation, our simulated
results decrease with $P_{b}$ when $P_{b}<60\,{\rm min}$, and then increase
with $P_{b}$.  So the changing trend of $\delta\left({\gamma}\right)$ is not
strictly consistent with the theoretical prediction, $\propto P_{b}^{4/3}$,
when $P_{b}<60\,{\rm min}$.  For DD model, according to {Eqs.\,(2.2b) and (2.2c) in \citet{Damour:1991rd},} the Einstein delay is always degenerate with the
R$\ddot{\rm o}$mer delay, and the degeneracy depends on the change of
relativistic advance of periastron \citep{Liu:2014uka}.  So we think that the
trend of $\delta\left({\gamma}\right)$ could come from the degeneracy between
the R$\ddot{\rm o}$mer delay and the Einstein delay.  For PSR-NS systems with
$P_{b} \sim 10\,{\rm min}$, they have a large $\dot{\omega}$, and considering a
4-yr observation, their periastrons have precessed several rounds.  Hence the
periodic change of $\omega$ can not eliminate degeneracy effectively enough and
leads to the trend of $\delta\left({\gamma}\right)$ different from the
theoretical prediction.  For binary systems with $P_{b} > 60\,{\rm min}$, the
change of $\omega$ is less than $2\pi$, so the trend of
$\delta\left({\gamma}\right)$ varying with $P_{b}$ is roughly consistent with
the theoretical prediction when considering different values of
$\omega_{0}$.\footnote{To verify the above arguments, we
simulate the fractional uncertainties of 5 PK parameters as functions of
$P_{b}$ by setting 1-yr observation and using the same observation cadence. In
this situation, the inflection point of trend change appears at 30 min.  It can
be understood that a short observation time $T_{\rm obs}$ gets a smaller change
for $\omega$ comparing with a 4-yr observation. In this case we have
$\Delta\omega < 2\pi$ for systems with $P_{b} > 30\,{\rm min}$, hence the
inflection point appears earlier.} So DD model can not avoid the degeneracy
because of the specific form of time delay terms, which makes $\gamma$  less
well measured when $\Delta{\omega}>2\pi$ in observation.}

We have showed the expected fractional uncertainties of PK parameters of
short-orbital-period PSR-NS systems which could be detected by LISA and SKA in
joint observation in the future.  But we need to emphasize that, strictly
speaking, the DD timing model is not sufficient to describe the orbital motion
of PSR-NS systems with a very short orbital period $P_b\sim$\,min, because it
only takes into account the leading PN order terms.  A PSR-NS system with
${P}_{b}=8\,{\rm min}$ is an extremely relativistic system, so, in principle, we
need to take high order effects into account in our simulation.  Since we mainly
use the fractional error of $\dot{P}_{b}$ to test gravity in the following, we
provide an extended analysis for $\dot{P}_{b}$.  Equation\,(\ref{eq:pbdot})
gives the 2.5\,PN formula for $\dot{P}_{b}$, $\dot{P}_{b}^{2.5\,{\rm PN}}$, and
we actually have appended it with the 3.5\,PN contribution \citep{1989Blanchet},
%------------------
\begin{align}
    \delta \dot{P}_{b}^{3.5\,{\rm PN}}&=\dot{P}_{b}^{2.5\,{\rm PN}}X^{3.5\,{\rm PN}} \,,
\end{align}
%------------------
where 
%------------------
\begin{align}
  X^{3.5\,{\rm PN}} = & \frac{\left(GmM_{\odot}\right)^{2/3}}{336\left(1-e^{2}\right)c^{2}f(e)}\left(\frac{P_{b}}{2\pi}\right)^{-2/3} \nonumber\\ 
  & \times{\bigg[}1273+\frac{16495}{2} e^{2} +\frac{42231}{8} e^{4}+\frac{3947}{16} e^{6}\nonumber \\
    & -\left(924+3381e^{2}+\frac{1659}{4}e^{4}-\frac{259}{4}e^{6}\right)\frac{m_{p}m_{c}}{m^{2}}\nonumber \\
    & +\left(3297e^{2}+4221e^{4}+\frac{2331}{8}e^{6}\right)\frac{m_{p}-m_{c}}{m}{\bigg]}\,.
\end{align}
%------------------
For a PSR-NS system with $P_{b}=8\,{\rm min},\,e=0.1$, one has $ X^{3.5\,{\rm
PN}} \sim 1\times10^{-4}$.  Our simulated fractional error of
$\dot{P}_{b}$ can reach $10^{-7}$ for $\sigma_{\rm TOA}=1\,\mu
{\rm s}$.  So, comparing with the simulated fractional error, the contribution
from $\delta \dot{P}_{b}^{3.5\,{\rm PN}}$ can be significant for an 8-min
system.  However $\delta \dot{P}_{b}^{3.5\,{\rm PN}}$ is still a small value
relative to $\dot{P}_{b}^{2.5\,{\rm PN}}$, so we use DD timing model to
simulate this situation which can give a correct estimate of magnitude of the
expected fractional uncertainties of PK parameters, and it can predict the
ability of a highly relativistic binary pulsar to test gravity.  We hope
that we can provide a more suitable timing model to simulate or process TOAs
for short-orbital-period PSR-NS systems in future work.

%---------------------------------------------------------------------
\section{Radiative tests of gravity}
\label{sec:test:GR}
%---------------------------------------------------------------------

In this section, we utilize the simulated results of $\dot{P}_{b}$ to do some
specific gravity tests.  Before we use the fractional error of $\dot{P}_{b}$
to test gravity, we need to emphasize that we should substract the
non-intrinsic contibution in $\dot{P}_{b}$ and use the intrinsic value
$\dot{P}_{b}^{\rm intr}$ in real gravity test.  The value of $\dot{P}_{b}$
from actual pulsar timing is $\dot{P}_{b}^{\rm obs}$ which contains the
dynamical contributions from the Shklovskii effect \citep{1970SvA....13..562S}
and the differential Galactic acceleration between the Solar System barycenter
and the pulsar system \citep{1991ApJ...366..501D,Lazaridis_2009}.  So, in
general, we need to subtract the contribution of the Shklovskii effect and the
Galactic acceleration difference in $\dot{P}_{b}^{\rm obs}$ first via,
%------------------
\begin{equation}\label{eq:pbobs_eq}
    \left(\frac{\dot{P}_{b}}{{P}_{b}}\right)^{\rm intr}=\left(\frac{\dot{P}_{b}}{{P}_{b}}\right)^{\rm obs}-\left(\frac{\dot{P}_{b}}{{P}_{b}}\right)^{\rm Shk}-
    \left(\frac{\dot{P}_{b}}{{P}_{b}}\right)^{\rm Gal}\,.
\end{equation}
%------------------
The second term of Eq.\,(\ref{eq:pbobs_eq}) is the
contribution from the Shklovskii effect which is due to the transverse motion
of the pulsar system \citep{1970SvA....13..562S},
{and the specific form is shown by Eq.\,(19) in \citet{Lazaridis_2009}.} 
The third term of Eq.\,(\ref{eq:pbobs_eq}) is the
contribution from the Galactic acceleration difference
\citep{1991ApJ...366..501D,Lazaridis_2009},
{and the specific form is given by Eq.\,(16) in \citet{Lazaridis_2009}.}

The actual fractional error of $\dot{P}_{b}^{\rm intr}$ depends on the errors of
$\dot{P}_{b}^{\rm Shk}$ and $\dot{P}_{b}^{\rm Gal}$. {So, the fractional
errors of $\dot{P}_{b}^{\rm Shk}$ and $\dot{P}_{b}^{\rm Gal}$ will affect
gravity tests. In other words, their fractional errors can determine how tight
the PSR-NS system can bound the parameter space of non-GR parameters.  For
$\dot{P}_{b}^{\rm Shk}$ and $\dot{P}_{b}^{\rm Gal}$, their fractional errors all
rely on the precision of $D$.}  At present, we use the Very Long Baseline
Interferometry (VLBI) or the dispersion measure (DM) to measure the distance of
binary pulsars.  For the nearby binary pulsar systems, VLBI can provide an
accurate measurement for $D$.  Once we have a measured distance with a small
error, the errors of $\dot{P}_{b}^{\rm Shk}$ and $\dot{P}_{b}^{\rm Gal}$ could
contribute little to the total error and the precision of $\dot{P}_{b}^{\rm
intr}$ would largely depend on $\dot{P}_{b}^{\rm obs}$, such as in the case of
PSRs J2222$-$0137 \citep{Cognard_2017} and J0737$-$3039 \citep{Kramer:2006nb}.
In the opposite case, for remote binary pulsar systems, we can not provide an
accurate measurement of $D$ by VLBI or DM.  The large uncertainty of $D$ results
in large errors of $\dot{P}_{b}^{\rm Shk}$ and $\dot{P}_{b}^{\rm Gal}$, which
restrict the precision of $\dot{P}_{b}^{\rm intr}$;
{for example, see PSR B1913+16 \citep{Weisberg_2016,Deller_2018}.}

So the accurate measurement of the distance of a binary pulsar system helps us
reduce the error contribution from astrophysical terms in $\dot P_b$.  Based
on a specific DNS system, \citet{Kyutoku:2018aai} provided an estimated
fractional uncertainty of distance for a 2-yr mission of LISA,
%------------------
\begin{align}\label{eq:deltaD}
    \frac{\Delta D}{D}=0.01\left(\frac{\rho}{200}\right)^{-1}\,.
\end{align}
%------------------
In \citet{Thrane:2019lwv}, considering a 4-yr mission of LISA, the fractional
uncertainty of distance can reach $0.006$ for a specific DNS system with
$\rho=360$. {So} LISA could improve the accuracy of pulsar
distance measurement significantly and then maybe avoid introducing large
errors caused by $\dot{P}_{b}^{\rm Shk}$ and $\dot{P}_{b}^{\rm Gal}$ (with a
well modelled Galactic potential).  Therefore, if LISA can detect a PSR-NS
system with a high SNR, it can improve the measure of distance of a PSR-NS
system.

To simplify the simulation, we did not include the possible error contribution
from $\dot{P}_{b}^{\rm Gal}$ and $\dot{P}_{b}^{\rm Shk}$, so we may derive {\it
optimistic} results when testing gravity.  In
Sec.\,\ref{sec:test:GR:constraint}, to be more complete, we provide a simple
calculation for the possible error contribution to $\dot{P}_{b}^{\rm intr}$ from
$\dot{P}_{b}^{\rm Shk}$ and $\dot{P}_{b}^{\rm Gal}$, and show the possible
impact on gravity tests.  We assume LISA can work well in distance measurement,
so we use a fractional error of $\Delta D/{D}=0.01$.  Actually, $D$ may not be
measured with such a high precision, {so even if the non-intrinsic effects
are considered, the results may still be optimistic because $D$ is assumed
to be measured with high precision.}

%---------------------------------------------------------------------
\subsection{Theoretical framework}
\label{sec:test:GR:theory}
%---------------------------------------------------------------------

Based on the expected measurement precision of the Five-hundred-meter Aperture
Spherical radio Telescope (FAST) and SKA, \citet{Liu:2014uka} simulated the
timing precision for black hole-pulsar (BH-PSR) binary systems. 
\citet{Seymour:2018bce} used the simulated parameter precision of
$\dot{P}_{b}$ from \citet{Liu:2014uka} to constrain specific gravity
theories.  In this section, following \citet{Seymour:2018bce}, we use the
simulated precision of $\dot{P}_{b}$ of the short-orbital-period PSR-NS
systems to constrain alternative theories of gravity.

In \citet{Seymour:2018bce}, the authors gave a generic formalism which
provides a mapping between generic non-GR parameters in the orbital decay rate
$\dot{P}_{b}$ and theoretical constants in various alternative theories of
gravity. Details can be found in Table\,I of \citet{Seymour:2018bce}.  A
parameterized non-GR modification to $\dot{P}_{b}$ reads,
%------------------
\begin{equation}
    \frac{\dot{P}_{b}}{{P}_{b}}=\frac{\dot{P}_{b}}{{P}_{b}}\bigg|_{\rm GR}\left(1+\Gamma v^{2n}\right)\,,\label{eq:mapping}
\end{equation}
%------------------
where $v$ is the relative velocity of two compact objects in a binary system,
$v=\left(\frac{2\pi M}{P_b}\frac{G}{c^{3}}\right)^{1/3}$,  and $M$ is the
total mass of binary system.  $\left({\dot{P}_{b}}/{{P}_{b}}\right)\big|_{\rm GR}$
is the quantity evaluated in GR [cf. Eq.\,(\ref{eq:pbdot})].  $\Gamma$ and $n$
are the non-GR parameters.  $\Gamma$ gives the overall magnitude of the
correction, and $n$ shows how the correction depends on $v$.  In
Eq.\,(\ref{eq:mapping}), a correction term proportional to $v^{2n}$
corresponds to a $n$-PN order correction to GR.  For different theories of
gravity, the non-GR parameter $\Gamma$ corresponds to different forms, which
are listed in Table\,I of \citet{Seymour:2018bce}, where a specific non-GR theory can be mapped to a set of $(\Gamma,n)$.  In our work, we use the same
mapping relation to test gravity.  We want to estimate how tightly the
short-orbital-period PSR-NS systems can constrain the parameters of non-GR
theories in the future.

%---------------------------------------------------------------------
\bgroup
\def\arraystretch{1.25}
\centering
\begin{figure}[htbp]
     \centering
     \includegraphics[width=0.48\textwidth]{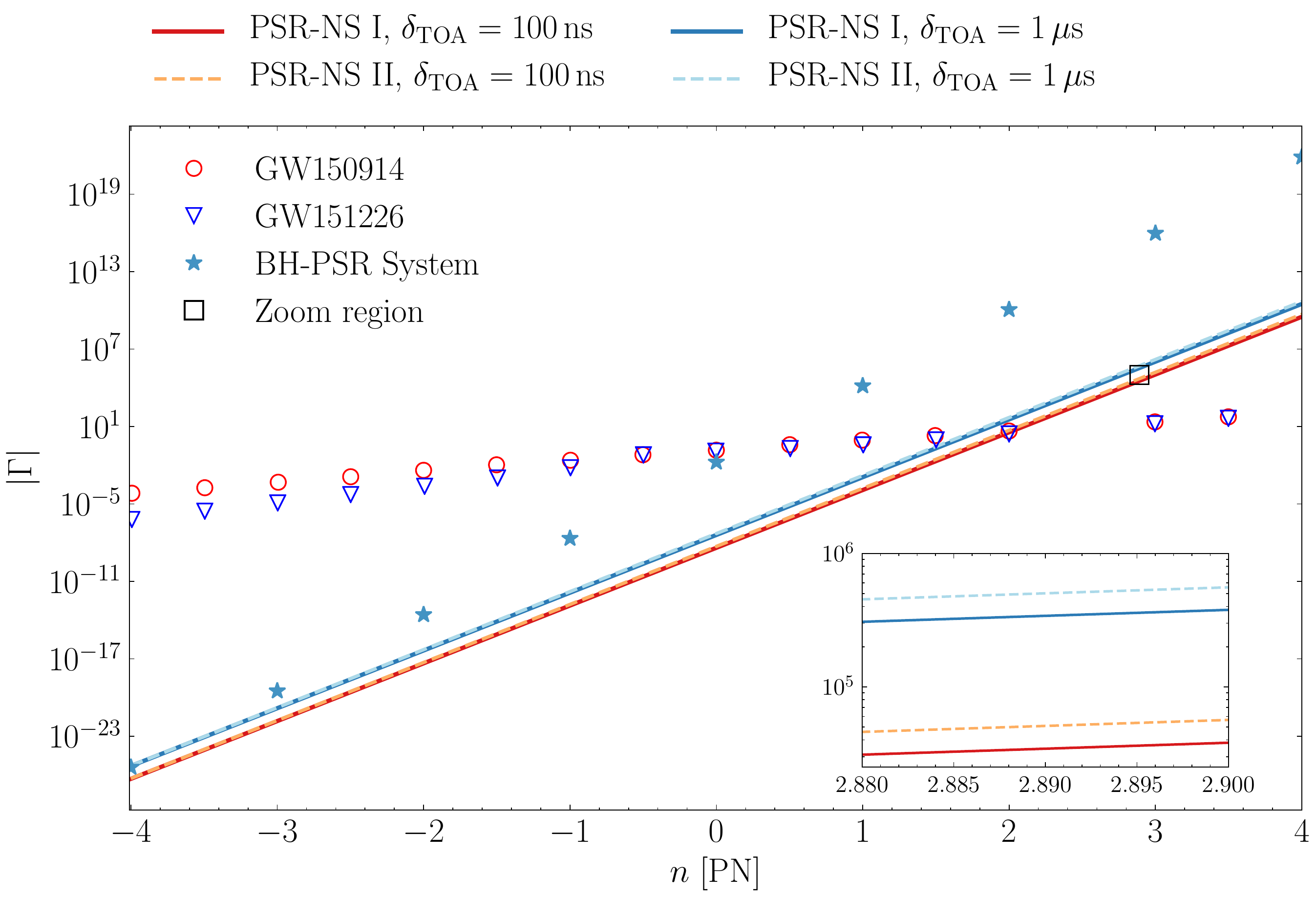}
     \caption{Upper bounds on $\Gamma$ as a function of $n$, where $n$ means
     that a correction term enters at $n$-PN order relative to GR. We present
     the predicted constraints with an 8-min PSR-NS system. We provide upper
     bounds from four sets of simulation: (i) PSR-NS I system with
     $\delta_{\rm TOA}=100\,{\rm ns}$ (red solid line),  (ii) PSR-NS I system
     with $\delta_{\rm TOA}=1\,{\rm \mu s}$ (blue solid line), (iii) PSR-NS II
     system with $\delta_{\rm TOA}=100\,{\rm ns}$ (light red dashed line), and
     (iv) PSR-NS II system with $\delta_{\rm TOA}=1\,{\rm \mu s}$ (light blue
     dashed line).  For comparison, we show bounds from a 3-day PSR-BH system
     \citep[blue stars; ][]{Seymour:2018bce}, and bounds from GW observations
     with GW150914 and GW151226 \citep[red circle and blue triangle;
     ][]{Yunes_2016}.  The insert plot is a zoom-in region of the
     black-squared region.  \label{fig:gammabound}}
\end{figure}
\egroup
%---------------------------------------------------------------------

Following {Eq.\,(4) in} \citet{Seymour:2018bce}, we define a fractional error $\delta$ for the observed ${\dot{P}_{b}}/{{P}_{b}}$,
{and we can use $\delta$ to bound $\Gamma$ by the following equation,}
\begin{equation}
    \left|\Gamma\right|<\frac{\delta}{v^{2n}}\,.
    \label{eq:Gamma}
\end{equation}
%------------------
In Sec.\,\ref{sec:sim}, we have simulated a series of fractional errors of
$\dot{P}_{b}$ for different $P_{b}$'s, and these fractional errors of
$\dot{P}_{b}$ correspond to the parameter $\delta$ here.  Hence we can use the
data to predict the constraints on $\Gamma$ for different PN orders.

We choose a PSR-NS system with an 8-min orbital period. 
Figure\,\ref{fig:gammabound} presents the predicted upper bounds on $\Gamma$ as
a function of PN orders.  For comparison, we also present the result from a
3-day orbital period PSR-BH system \citep{Seymour:2018bce}.  For a PSR-NS system
with an 8-min orbital period, {its relative velocity $v$ approximately
equals to $5.7\times 10^{-3}$ which is 5 times larger than the velocity of a
3-day orbital period PSR-BH system, whose $v\sim 1\times 10^{-3}$.  For negative
PN, $\Gamma\propto v^{|2n|}$, so the larger value of $n$,  a binary system with
a larger $v$ is more disadvantageous for bounding $\Gamma$. For example, if
$n=-4$, for an 8-min PSR-NS system, its value of $v^{|2n|}$ is $4\times10^{5}$
times larger than that of a 3-day PSR-BH system, which greatly reduces the
restriction on $\Gamma$.}  However, the 8-min orbital period PSR-NS system has a
smaller fractional error $\delta$, that makes it possible to provide a stronger
constraint on $\Gamma$.  Combining those two factors, the constraint from an
8-min PSR-NS system is tighter than the constraint from a 3-day PSR-BH system at
some negative PN orders, as shown in Fig.\,\ref{fig:gammabound}.  At positive PN
orders, a larger $v$ is more competitive, so an 8-min PSR-NS system can
naturally provide a tighter constraint than a 3-day PSR-BH system.  Like
\citet{Seymour:2018bce}, we also show the limits from GW observations  in
Fig.\,\ref{fig:gammabound}, using GW150914 and GW151226 as examples
\citep{Yunes_2016}.  Because GWs detected by LIGO/Virgo come from the phase of
merger, which indicates a larger relative velocity, it leads to a tighter
restriction on $\Gamma$ at positive PN orders.  But due to the small $\delta$ of
the 8-min PSR-NS system, it can give a slightly tighter restriction than GW
observations at positive PN order when $n\lesssim 2$. 

Overall, the results in Fig.\,\ref{fig:gammabound} show that the 8-min PSR-NS
system can provide a tight bound for the non-GR theories of negative PN orders
when $n \gtrsim -4$.  However, when comparing these results in a strict way, one
should be aware that the prediction may depend on the details of specific
theories. NSs and BHs can behave quite differently in alternative gravity
theories. We leave this to a future detailed study.  In any case, the system
that we considered here will provide very useful and complementary constraints.
We underline that here the corrections of different PN orders are relative to
GR's quadrupole radiation, not to the equations of motion.

%---------------------------------------------------------------------
\subsection{Projected constraints}
\label{sec:test:GR:constraint}
%---------------------------------------------------------------------

In Fig.\,\ref{fig:gammabound}, we show the constraint from an 8-min PSR-NS
system on parameter $\Gamma$ as a function of PN orders.  In
Sec.\,\ref{sec:test:GR:theory}, the generic formalism (\ref{eq:mapping})
offers a mapping between generic non-GR parameters in the orbital decay rate
and theoretical parameters in various alternative theories of gravity
\citep{Seymour:2018bce}.  We can use this mapping to restrict specific
theories of gravity, and then predict how do the short-orbital-period PSR-NS
systems constrain these theories.

%---------------------------------------------------------------------
\begin{figure*}[htbp]
    \centering
    \includegraphics[width=0.9\textwidth]{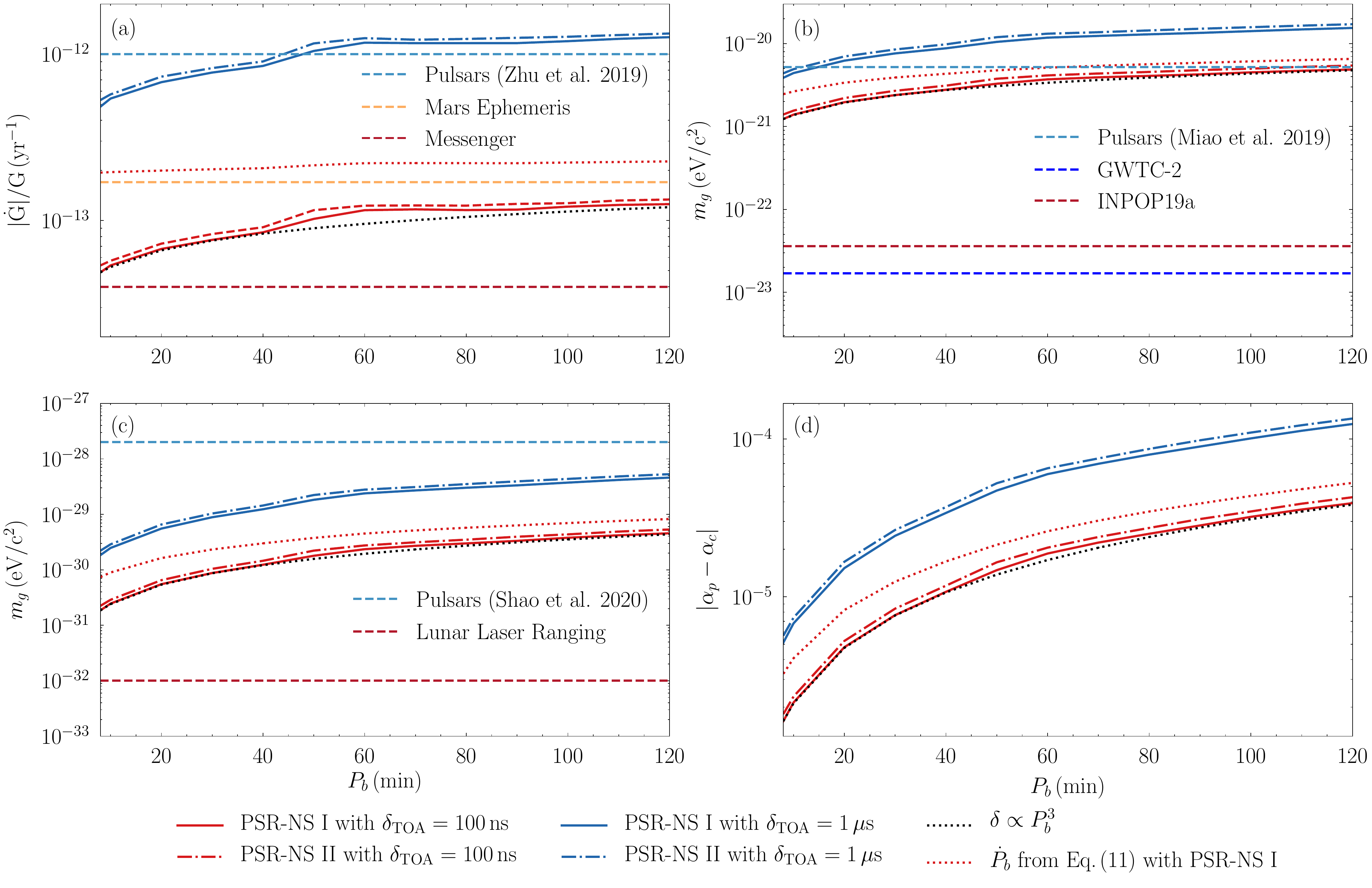}
    \caption{Projected constraints on non-GR parameters with short-orbital-period
    PSR-NS systems using the SKA. We provide upper bounds from four sets of
    simulation: (i) PSR-NS I system with $\delta_{\rm TOA}=100\,{\rm ns}$ (red
    solid line), (ii) PSR-NS I system with $\delta_{\rm TOA}=1\,{\rm \mu s}$ (blue
    solid line), (iii) PSR-NS II system with $\delta_{\rm TOA}=100\,{\rm ns}$ (red
    dotted-dashed line), and (iv) PSR-NS II system with $\delta_{\rm TOA}=1\,{\rm
    \mu s}$ (blue dotted-dashed line).  The black dotted line is the trend of
    bound predicted from theory, i.e., $\delta \propto P_{b}^{3}$
    \citep{Damour:1991rd,2004hpa..book.....L}.  The red dotted line is the
    constraint on non-GR parameters from fractional error of $\dot{P}_{b}^{\rm
    intr}$ (using PSR-NS I with $\delta_{\rm TOA}=100\,{\rm ns}$ for an
    illustration) which has taken into account error contributions from
    $\dot{P}_{b}^{\rm Shk}$ and $\dot{P}_{b}^{\rm Gal}$.  Panel (a) shows 
    constraints on $|\dot{G}|/G$ as a function of $P_{b}$ in varying-$G$ theory.
    For comparison, we show bounds from Solar System:  the bound with Mars
    Ephemeris \citep{Will_2014} in a gold dashed line, and the bound with NASA
    Messenger \citep{2018_MESSENGER} in a red dashed line;  we also show the
    constraint from binary pulsars with a blue dashed line \citep{Zhu_2018}.
    Panel (b) shows constraints on graviton mass $m_{g}$ in a Fierz–Pauli-like
    massive gravity theory.  For comparison, the constraint on $m_{g}$ from Solar
    System (marked ``{INPOP19a}'') is shown in a red dashed line
    \citep{Bernus_2020}; the constraint from GW detection (marked ``GWTC-2'')
    is shown in a blue dashed line
    \citep{LIGOScientific:2020tif}; the constraint from
    combination of existing pulsar observations is given in a cyan dashed line
    \citep{Miao:2019nhf}.   Panel (c) shows constraints on graviton mass $m_{g}$
    in the cubic Galileon massive gravity theory.  The latest bound from binary
    pulsars is shown with a light blue dashed line \citep{Shao_2020}.  The best
    bound is from lunar laser ranging experiment \citep[red dashed line;
    ][]{moon_2003}.  Panel (d) shows constraints on
    $\left|\alpha_{p}-\alpha_{c}\right|$ in DEF gravity theory.
    \label{fig:gravitytest}}
    \end{figure*}
    %---------------------------------------------------------------------

%---------------------------------------------------------------------
\subsubsection{Varying-$G$ theory}
\label{sec:test:GR:constraint:vary_g}

In GR, the Newtonian gravitational constant $G$ does not change with time. 
But in some alternative theories, the locally measured Newtonian gravitational
constant $G$ may vary with time, because of the evolution of the Universe
\citep{Will_2014}.  The gravitational constant which varies with time violates
local position invariance, that leads a violation of strong equivalence
principle (SEP).  Hence proving whether the gravitational constant varies with
time helps us test the SEP.  If the gravitational constant is changing with
time, it can cause changes in the compactness and Arnowitt-Deser-Misner (ADM)
mass of a compact star, which would result in an additional contribution to
the orbital period decay rate $\dot{P}_{b}$ \citep{PhysRevLett.65.953}.

The relation between $\dot{P}_{b}$ and $\dot{G}$ is given by \citet{PhysRevLett.65.953}, 
%------------------
\begin{equation}
    \frac{\dot{P}_{b}^{\dot{G}}}{{P}_{b}}=-2\frac{\dot{G}}{G}\left[1-\frac{2m+m_{c}}{2m}s_{p}-\frac{2m+m_{p}}{2m}s_{c}\right]\,,
    \label{eq:gdot}
\end{equation}
%------------------
where $s_{i}$ denotes the ``sensitivity'' of the NS, which depends on the mass and the EOS of NS,
%------------------
\begin{equation}
    s_{i}\equiv-\frac{\partial \ln{M_{i}}}{\partial\ln{G}},\quad\quad (i=p,c)\,,
\end{equation}
%------------------
where $M_{i}=m_{i}M_{\odot}$ (notice that in this paper $m_i$ is the mass in
unit of Solar mass). In GR, $s_{i}=0$.  We use the value of $s_{i}$ as per
\citet{Damour_1992}, \footnote{{ In principle, this relation is based on
the Jordan–Fierz–Brans–Dicke gravity and the EOS AP4 \citep{Lattimer_2001}. However, a slightly
different value will not affect our results at the first-order approximation.}}
%------------------
\begin{equation}
    s_{i}=0.16\,\left(\frac{M_{i}}{1.33M_{\odot}}\right)\,.
\end{equation}
%------------------
With Eqs.\,(\ref{eq:Gamma}--\ref{eq:gdot}), we get the corresponding
expression of $\Gamma$,
%------------------
\begin{equation}
    \Gamma=\frac{5}{48}\frac{\dot{G}}{c^{3}}\frac{m^{3}M_{\odot}}{m_{p}m_{c}}\left[1-\frac{2m+m_{c}}{2m}s_{p}-\frac{2m+m_{p}}{2m}s_{c}\right]\frac{1}{f(e)}\,,
    \label{eq:gammagdot}
\end{equation}
%------------------
and it is a $-$4 PN order correction relative to GR.  So combining
Eq.\,(\ref{eq:mapping}) and Eq.\,(\ref{eq:gammagdot}), we get the upper bound
on $\dot{G}$ by,
%------------------
\begin{equation}
    \frac{|\dot{G}|}{G}<-\frac{\delta}{2} \left( \frac{\dot{P_{b}}}{P_{b}} \right)_{\rm GR} {\left(1-\frac{2m+m_{c}}{2m}s_{p}-\frac{2m+m_{p}}{2m}s_{c}\right)^{-1}}\,.
\end{equation}
%------------------

There is an associated dipole radiation of GWs for most gravitational theories
that violate SEP.  The dipole radiation of GWs can contribute an extra
$\dot{P}_{b}^{\rm dipole}$ which is degenerate with $\dot{P}_{b}^{\dot{G}}$.
The degeneracy indicates that we need to use multiple binary pulsar systems to
break the degeneracy.  In our work, for
the sake of simplicity, we ignore the possible degeneracy and consider only
one violation to GR at a time.

Figure\,\ref{fig:gravitytest}\,(a) shows the predicted upper limit on
$\dot{G}$ from short-orbital-period PSR-NS systems with
$P_{b}\in[8,\,120]\,{\rm min}$ to be detected by the SKA.  The best bound is
$|\dot{G}|/G<4.8\times10^{-14}\,{\rm yr}^{-1}$ which is from the 8-min PSR-NS
I system with $\sigma_{\rm TOA} =100\,{\rm ns}$. 

The simulated fractional uncertainty of $\dot{P}_{b}$ varies with $P_{b}$, 
that is largely consistent with what is theoretically predicted \citep[namely
$\delta \propto P_{b}^{3}$; ][]{Damour:1991rd,2004hpa..book.....L}.  Because
of the degeneracy of some delay terms, there are still slight differences
between theoretic prediction and our simulation.  In
Fig.\,\ref{fig:gravitytest}\,(a), the black dotted line is the trend by
theoretical prediction, $\delta \propto P_{b}^{3}$, and we use PSR-BH I system
with $\sigma_{\rm TOA} =100\,{\rm ns}$ to illustrate.
{Because we show the
constraint on $|\dot{G}|/G$, which is a cumulative effect in one year, so there
is a noticeable deviation in $P_{b}\in[40,\,60]\,{\rm min}$.}

We do not plot the bounds of a PSR-BH system from \citet{Seymour:2018bce}, and
their results are similar to ours.  Although the 8-min orbital period system
has a very high precision for $\dot{P}_{b}$, as mentioned before, the varying
$G$ theory is a $-$4 PN order correction relative to GR.  A system with a
large $v$ is disadvantageous to provide a tighter bound on $\dot{G}$, so the
two systems (PSR-NS and PSR-BH) have similar results.

To provide a more complete analysis, we also consider the fractional error
which contains error contributions from $\dot{P}_{b}^{\rm Shk}$ and
$\dot{P}_{b}^{\rm Gal}$ to restrict $\dot{G}$ [see Eq.~(\ref{eq:pbobs_eq})].
We assume that LISA can help us determine the distance of
the PSR-NS system very well.  So for the fractional error of $D$, we take
$0.01$.\footnote{{The parameters $\mu_{\alpha}$, $\mu_{\delta}$, $l$, $b$ and $D$ in $\dot{P}_{b}^{\rm Shk}$ and $\dot{P}_{b}^{\rm Gal}$ come from PSR J0737$-$3039 \citep{Kramer:2006nb, Hu:2020ubl}.}}  Considering the error
contributions from $\dot{P}_{b}^{\rm Shk}$ and $\dot{P}_{b}^{\rm Gal}$, we
show in Fig.\,\ref{fig:gravitytest}\,(a) the bound from PSR-NS I systems with
$\delta_{\rm TOA}=100\,{\rm ns}$ on $\dot{G}$ with the red dotted line.  We find
that by introducing the errors of $\dot{P}_{b}^{\rm Shk}$ and
$\dot{P}_{b}^{\rm Gal}$, our results become worse.  Although our distance
accuracy is high, the precision of other parameters that we {adopt from
PSR J0737$-$3039} are low, which prevent us from
limiting the non-GR parameters to high precision.  
If we, likely with the
SKA, further improve the measurement accuracy of these parameters in the
future, we will constrain $\dot{G}$ better.

In order to compare, we show the current bounds from the Solar System
experiments in Fig.\,\ref{fig:gravitytest}\,(a).  We notice that the estimated
bound of an 8-min PSR-NS system is slightly weaker than the Solar System result
from NASA Messenger \citep[red dashed line; ][]{2018_MESSENGER}.  For now, the
best constraint comes from the Solar System experiment, which is from weak
field.  However, the sensitivity $s_{i}$ is a body-dependent quantity, so it
could lead to an enhance of $\dot{G}$ for strongly self-gravitating bodies.
NSs are strongly self-gravitating bodies, so it provides the constraint on
$\dot{G}$ from a strong field.  The constraint on $\dot{G}$ from PSR-NS systems
can provide a complementary constraint compared with the constraint from the
Solar System experiment.  {One point that needs to be emphasized is that we
can improve the precision of the parameters through long-term observation.  For
$\dot{P}_{b}$, the expected dependence of fractional uncertainty on observing
time, $T_{\rm obs}$, is proportional to $T_{\rm obs}^{-5/2}$.  Therefore,
according to the theoretical prediction, compared with a 4-yr observation, the
measurement precision of $\dot{P}_{b}$ will increase by about 10 times {for a 10-yr timing}.  We
simulate a 10-yr timing for PSR-NS I system with $\delta_{\rm TOA}=100\,{\rm
ns}$ by using the same cadence, and the result is consistent with the
theoretical prediction; that is, the measurement precision is improved by an
order of magnitude.  Hence, a 10-yr observation could increase the projected
limit to $|\dot{G}|/G<5\times10^{-15}\,{\rm yr}^{-1}$ for an 8-min PSR-NS I
system with $\delta_{\rm TOA}=100\,{\rm ns}$.  Thus, an 8-min PSR-NS system has
the potential to give tighter bound than the present result from the Solar
System for $\dot{G}$.}

We also provide the
previous upper limit on $\dot{G}$ from binary pulsar systems, that combined
the results from PSRs J0437$-$4715, J1738+0333, and J1713+0747 \citep[blue
dashed line; ][]{Zhu_2018}.  Our simulated results show that the constraint on
$\dot{G}$ in strong field can be improved significantly.

%---------------------------------------------------------------------
\subsubsection{Fierz–Pauli-like massive gravity}
\label{sec:test:GR:constraint:lv-massive}

In GR, gravitation is mediated by a massless spin-2 graviton, and graviton is
massless in most theories of gravity.  The existence of graviton is not
experimentally confirmed yet.  
There are some theories of gravity where
graviton has mass {and they cannot be
completely ruled out yet.}
{So we can base on different theories to bound graviton mass}
\citep{de_Rham_2017},
and we show some results in Fig.\,\ref{fig:gravitytest}\,(b).  If graviton is
massive, the gravitational potential possesses an exponential Yukawa
suppression.  In the Solar System, which can be viewed as a static field, the
Yukawa potential can cause planets to behave differently from the Newtonian
potential which is $\propto 1/r$.  \citet{Bernus_2020} used the planetary ephemeris
{INPOP19a} to constrain graviton mass and provided {$m_{g}<3.62\times10\,{\rm
eV}/c^{2}$}, shown with a red dashed line in Fig.\,\ref{fig:gravitytest}\,(b). 
The constraint on $m_{g}$ from the Solar System is a bound in weak field; we
also provide two constraints from strong field.  The first bound comes from GW
data.  If gravity is propagated by a massive field, the massive graviton can
lead to a modified dispersion relation.  In other words, the velocity of the
graviton depends on the GW frequency.  Using the modified dispersion relation,
the LIGO/Virgo Collaboration bound the graviton mass to be
$m_{g}<1.8\times10^{-23}\,{\rm eV}/c^{2}$ by combining the data of 31 GW
events from the catalog GWTC-2
\citep{LIGOScientific:2020tif}.  This result is represented
by a blue dashed line in Fig.\,\ref{fig:gravitytest}\,(b).  This bound coming
from the direct detection of GWs is an effect of GWs' phase accumulation.  The
second bound is from binary pulsar systems, which is based on a
Fierz–Pauli-like massive gravity.  \citet{Finn_2002} presented a
phenomenological term in the Lagrangian for linearized gravity with a mass
term $\sim m_{g}^{2}\left(h_{\mu\nu}^{2}-\frac{1}{2}h^{2}\right)$ which is
similar to the Fierz–Pauli mass term $\sim
m_{g}^{2}\left(h_{\mu\nu}^{2}-h^{2}\right)$ \citep{Fierz:1939ix}, but with a different coefficient
for $h^{2}$.  The choice of this linear mass term ensures that the theory
reduces to GR when $m_{g}\rightarrow0$ {[avoiding the van Dam-Veltman–Zakharov (vDVZ) discontinuity]} and the derived wave equation takes the
standard form \citep{Finn_2002},
%------------------
\begin{align}
    \left( \square - m_g^2 \right) \bar h_{\mu\nu} + 16\pi T_{\mu\nu} = 0 \,,
    \label{eq:finnfield}
\end{align}
%------------------
with an $h$-independent source $T_{\mu\nu}$.  We have defined $\bar h_{\mu\nu}
\equiv h_{\mu\nu} - \frac{1}{2} \eta_{\mu\nu} h$.  \citet{Finn_2002} worked
out {the relation between $m_{g}$ and $\delta$,}
%------------------
\begin{align}
   m_{g}^{2} <
   \frac{24}{5}F(e)\left(\frac{2\pi\hbar}{c^{2}P_{b}}\right)^{2}\delta\,,
   \label{eq:finnmassbound}
\end{align}
%------------------
{where $F(e)=f(e)\left(1-e^{2}\right)^{1/2}$ [see Eq.~(\ref{eq:fe}) for $f(e)$], and $\hbar$ is the reduced Planck
constant.}
With Eq.\,(\ref{eq:finnmassbound}), we can use binary pulsars to restrict
$m_{g}$.  \citet{Finn_2002} provided a method to constrain graviton mass in a
dynamic regime.  The Fierz–Pauli-like massive gravity gives a $-$3 PN order
correction relative to GR.  By comparison with Eq.\,(\ref{eq:Gamma}), we can
obtain the form of $\Gamma$,
%------------------
\begin{align}
    \Gamma < \frac{5}{24}m^{2}_{g}m^{2}\frac{M_{\odot}^{2}G^{2}}{c^{2}\hbar^{2}}\frac{1}{F(e)}\,.
    \label{eq:gammafinn}
\end{align}
%------------------
\citet{Finn_2002} used PSRs B1913+16 and B1534+12 to bound $m_{g}$.  Recently,
using 9 well-timed binary pulsars and the Bayesian theorem,
\citet{Miao:2019nhf} provided an improved bound, $m_{g}<5.2\times10^{-21}\,
{\rm eV}/c^{2}$, which is shown with a light blue dashed line in
Fig.\,\ref{fig:gravitytest}\,(b). 

Here, we use the simulated $\delta$ of $\dot{P}_{b}$ of the
short-orbital-period PSR-NS systems by the SKA, to predict the constraints on
graviton mass.  The results of bounds are shown in
Fig.\,\ref{fig:gravitytest}\,(b).  The results show that the PSR-NS system with
an 8-min short orbital period can give a tighter limit than previous results in
\citet{Miao:2019nhf} by roughly a factor of 4, {$m_{g}<1.2\times10^{-21}\,
{\rm eV}/c^{2}$.} Although the projected bounds from PSR-NS systems are looser
than the bounds from GWs and the Solar System, they reflect a dynamical process
of two compact objects and a strong field effect. {The constraint from GWs
is based on a modified dispersion relation which is an effect of accumulation
in GW phases, so it is a kinematic effect from propagation.  The constraint
from the Solar System reflects an effect from a static weak field. Hence the
restrictions from the binary pulsars are complementary to these constraints.}
{For an 8-min PSR-NS I system with $\delta_{\rm TOA}=100\,{\rm ns}$,
simulated results for a 10-yr observation span improve it to
$m_{g}<4.0\times10^{-22}\, {\rm eV}/c^{2}$. Because $m_{g}\propto\delta^{1/2}$,
the increase in precision of $\dot{P}_{b}$ can not improve the limit on $m_{g}$
significantly in this case.} We also show the trend of theoretical prediction,
$\delta \propto P_{b}^{3}$, by the black dotted line in
Fig.\,\ref{fig:gravitytest}\,(b).  Our results do not significantly deviate
from the theoretical trend. In addition, we give the restriction on $m_{g}$
when there is error contribution of $\dot{P}_{b}^{\rm Shk}$ and
$\dot{P}_{b}^{\rm Gal}$, and it is shown with a red dotted line in
Fig.\,\ref{fig:gravitytest}\,(b).

\subsubsection{Cubic Galileon massive gravity}
\label{sec:test:GR:constraint:cubicgalileon}

If graviton is massive, there may be extra scalar and vector degrees of
freedom.  The vector degrees of freedom can be ignored, because they usually
do not couple to matters in the decoupling limit \citep{de_Rham_2017}.  The
extra scalar degree of freedom can couple to matters which leads to a fifth
force, and the fifth force would make theory of massive gravity fail to
recover GR when $m_{g}\rightarrow0$ with the so-called vDVZ discontinuity. {However the} vDVZ discontinuity {can be solved} by introducing screening mechanisms,
e.g. the Vainshtein mechanism \citep{VAINSHTEIN1972393}. The scalar degree of freedom becomes strongly-coupled and is strongly suppressed, {because of Vainshtein mechanism,} so the
deviation from GR is suppressed.  In other words, the fifth force is
suppressed and the metric field propagates the gravitational force.

The simplest model used to display Vainshtein mechanism is called the cubic Galileon \citep{Luty_2003} which is a Lorentz invariant massive gravity model.
The massive graviton causes extra GW radiations, 
and there are not only quadrupole radiation but also monopole and dipole radiations that need to be considered.
\citet{de_Rham_2013} calculated the contributions of these three radiations, 
and found that the dipole radiation is orders of magnitude smaller than the other two types and the quadrupole radiation has the largest contribution in the vast majority of cases.
In fact, we calculated the contribution of monopole radiation,
and found that it can be ignored indeed.
So we just consider quadrupole radiation in our work. 

The extra contribution to $\dot{P}_{b}$ from the quadrupole radiation power in
the cubic Galileon model is \citep{de_Rham_2013},
\begin{align}
  \dot{P}_{b}^{\rm quad}=-\frac{15}{\sqrt{2}}\lambda^{2}\pi^{3/2}\frac{(GM)^{5/6}}{n_{b}^{1/6}\hbar c^{1/2}}\frac{m_{p}m_{c}\mathcal{Y}^{2}}{m^{2}}m_{g}f_{\rm cg}(e)\,,
   \label{eq:galileonpbdot}
\end{align}
where {$\mathcal{Y} \equiv \big(m_{p}^{1/2}+m_{c}^{1/2}\big){\big/}m^{1/2}$, the constant $\lambda=0.2125$ \citep{Shao_2020}, and}
$f_{\rm cg}(e)=\sum_{n=0}^{\infty}\left|I_{n}^{\rm quad}(e)\right|^{2}$,
%-------------------------------------------
\begin{align}
  I_{n}^{\rm quad}(e)=\frac{\left(1-e^{2}\right)^{3/2}}{2\pi}n^{7/4}\int^{2\pi}_{0}\frac{\exp\left[i(2-n)x\right]}{\left(1+e\cos{x}\right)^{3/2}} {\rm d} x\,.
\end{align}
%-------------------------------------------
{We choose $n=30$ in our calculation,
which is enough for $e=0.1$~\citep{Shao_2020}.}
With Eq.\,(\ref{eq:mapping}) and Eq.\,(\ref{eq:galileonpbdot}), we can provide
the constraint of $m_{g}$,
\begin{align}
  m_{g}<\frac{8}{5\lambda^{2}}\frac{1}{f_{\rm cg}(e)}\frac{m^{1/2}M_{\rm Pl}}{M^{2}_{\rm quad}}\frac{\mathcal{P}_{\rm GR}}{n_{b}^{1/2}(n_{b}a)^{3}}\left(\hbar M_{\odot}\right)^{1/2}\delta\,,
   \label{eq:galileon_mg}
\end{align}
where {$n_{b}=2\pi/P_{b}$ is the orbital angular frequency, $a$ is the
semi-major axis, $M_{\rm Pl}\equiv \left(\hbar
c/8\pi G\right)^{1/2}$ is the reduced Planck mass,}
$\mathcal{P}_{\rm GR}$ is the radiation power of GR,
\begin{align}
 \mathcal{P}_{\rm GR}=\frac{32m}{5}\frac{m^{2}_{p}m^{2}_{c}}{a^{5}}\frac{G^{4}M_{\odot}^{5}}{c^{5}}f(e)\,,
   \label{eq:grradiation}
\end{align}
and,
%------------------
\begin{align}
  M_{\rm quad} \equiv \frac{m_{p}m_{c}}{m}\mathcal{Y}M_{\odot}\,.
\end{align}
%------------------
The cubic Galileon massive gravity model provides a $-\frac{11}{4}$ PN order
correction relative to GR, and $\Gamma$ is
\begin{align}
 \Gamma=\frac{25\lambda^{2}}{2^{8}}\frac{M^{2}_{\rm quad}}{M_{\rm Pl}}\frac{m^{3}m_{g}}{m_{p}^{2}m_{c}^{2}}\left(\frac{G}{\hbar c}\right)^{1/2}\frac{f_{\rm cg}(e)}{f(e)}\frac{1}{M_{\odot}}\,.
   \label{eq:gcgamma}
\end{align}
With Eq.\,(\ref{eq:galileon_mg}), we show constraints from our simulated
results in Fig.\,\ref{fig:gravitytest}\,(c).  The light blue dashed line is
the bound from a combination of binary pulsar systems, which gives
$m_{g}<2.0\times10^{-28}\,{\rm eV}/c^{2}$ \citep{Shao_2020}.  Compared with
it, the short-orbital-period PSR-NS systems can greatly improve the constraint
on $m_{g}$, and a projected bound from an 8-min PSR-NS system is
$m_{g}<1.9\times10^{-31}\,{\rm eV}/c^{2}$.  Because $m_{g}\propto\delta$, the
deviation with respect to the trend predicted by theory (black dotted line) is
more significant than that of the Fierz–Pauli-like massive gravity.  The red
dotted line in Fig.\,\ref{fig:gravitytest}\,(c) is the restriction which
considers the possible error contributions from $\dot{P}_{b}^{\rm Shk}$ and
$\dot{P}_{b}^{\rm Gal}$.  In Fig.\,\ref{fig:gravitytest}\,(c), the best
restriction is from the red dashed line, $m_{g}<10^{-32}\,{\rm eV}/c^{2}$
\citep{moon_2003}.  It is a bound from the lunar laser ranging experiments in
the weak field. 
{For a 10-yr observation, an 8-min PSR-NS I system with $\delta_{\rm TOA}=100\,{\rm ns}$ can provide a bound $m_{g}<2.0\times10^{-32}\,{\rm eV}/c^{2}$, which is comparable to the lunar laser ranging experiments.
So if we can observe these systems for decades, we could provide a tighter constraint than the current result.}
{So the} short-orbital-period PSR-NS systems {have potential to significantly} improve the
restriction on $m_{g}$ in the strong field.

%-----------------------------------------------------------------------------
\subsubsection{Scalar-tensor theory}
\label{sec:test:GR:constraint:scalar-tensor}
%-----------------------------------------------------------------------------

In GR, gravity is mediated only by a long-range, spin-2 tensor field,
$g_{\mu\nu}$, but in alternative theories there can be scalar or vector fields
involved in the gravitational interaction.  The scalar-tensor gravity, which
adds a spin-0 scalar field, $\varphi$, is the most natural alternative theory.
In our work, we take Damour and Esposito-Far$\grave{\rm e}$se (DEF) theory  as
an example \citep{Damour_1992,Damour:1993hw,Damour_1996}.  In the weak field,
the DEF gravity theory can be reduced to GR, which means that it can pass the Solar
System tests.  However, in the strong field, DEF theory can be completely
different from GR, which is due to a strong field \textit{spontaneous
scalarization} phenomenon \citep{Damour:1993hw,Damour_1996}, so binary pulsar
systems provide an ideal laboratory to study the DEF theory.

In DEF theory, the GW damping is different from GR.  There is an additional
contribution of GW radiation from the scalar field, which leads to an extra
orbital decay rate $\dot{P}_{b}^{\rm scalar}$, and it contains monopole,
dipole, and quadrupole contributions.  We only consider
the dipole contribution which is a $-$1 PN correction relative to GR, and the
expression of $\dot{P}_{b}^{\rm {dipole}}$ is
\citep{Damour:1993hw,Damour_1996},
\begin{align}\label{eq:dipole}
     \dot{P}_{b}^{\rm {dipole }}&=-2 \pi T^{*}_{\odot}d(e)\left(\frac{2 \pi}{P_{b}}\right) \frac{m_{p} m_{c}}{m_{p}+m_{c}}\left(\alpha_{p}-\alpha_{c}\right)^{2}\,,
\end{align}
where {$\alpha_{i}$ is the coupling strength between the scalar field and matter,} $d(e) \equiv
\left({1+\frac{1}{2}e^{2}}\right){\left(1-e^{2}\right)^{-5/2}}$,
$T^{*}_{\odot}=G^{*}M_{\odot}/c^{3}$, 
$G^{*}=G/\left(1+\alpha_{0}^{2}\right)${\footnote{The ``$*$'' denotes the Einstein frame.}, and $\alpha_{0}$ is the linear coupling of
scalar field to matter fields}. For simplification, we ignore the
difference between $G^{*}$ and $G$. {Because the current weak-field
constraint on $\alpha_{0}^{2}$ has reached $10^{-5}$
\citep{2003Natur_Bertotti}, we set $G^{*}=G$.}  The effective scalar couplings
$\alpha_{p}$ and $\alpha_{c}$ depend on the specific EOS of NSs.

In our work, we only use the simulated fractional error to constrain
$\left|\alpha_{p}-\alpha_{c}\right|$, and we do not further analyze
$\alpha_{p}$ by using specific EOSs.\footnote{The method that uses binary
pulsar systems to bound $\alpha_{p}$ with various EOSs can refer to, {e.g., \citet{Shao:2017gwu},
\citet{Zhao:2019suc}, and \citet{Guo:2021leu}}.} The spontaneous scalarization of DEF theory happens
in strong field, so it is conducive to test DEF theory in PSR-NS systems.
With Eq.\,(\ref{eq:mapping}) and Eq.\,(\ref{eq:dipole}), we can provide the
constraint of $\left|\alpha_{p}-\alpha_{c}\right|$, 
\begin{equation}
\left(\alpha_{p}-\alpha_{c}\right)^{2}=\frac{96}{5}T_{\odot}^{2/3}\left(\frac{P_{b}}{2\pi}\right)^{-2/3}D(e)\left(m_{p}+m_{c}\right)^{2/3}\delta\,,\label{eq:deltalpha}
\end{equation} 
where $D(e) \equiv f(e)/d(e)$.  Using Eq.\,(\ref{eq:deltalpha}), we provide the
results in Fig.\,\ref{fig:gravitytest}\,(d).  {We notice that the best
constraint on $\left|\alpha_{p}-\alpha_{c}\right|$ can reach
$1.6\times10^{-6}$.  In the work of \citet{Shao:2016ezh}, they showed
constraints $\sim10^{-3}$ for several best measured PSR-WD systems where
$\left|\alpha_{p}-\alpha_{c}\right|=\left|\alpha_{p}-\alpha_{0}\right|$ due to
the companion being a WD.  \footnote{{If the companion is a WD,
$\alpha_{c}\simeq\alpha_{0}$.}} The simulated 8-min PSR-NS system seems to
provide a tighter bound on $\left|\alpha_{p}-\alpha_{c}\right|$, but the
companion of system is a NS, so $\alpha_{c}$ could be very close to
$\alpha_{p}$ which makes $\left|\alpha_{p}-\alpha_{c}\right|$ small for most
cases.   Our predicted constraints on $\left|\alpha_{p}-\alpha_{c}\right|$ come
from a  simplified analysis, and the real situation needs to be discussed with
EOS dependence included.} In the figure, we also provide the trend of
prediction from theory (black dotted line) and the results containing the
possible error contributions from $\dot{P}_{b}^{\rm Shk}$ and $\dot{P}_{b}^{\rm
Gal}$ (red dotted line).

%-----------------------------------------------------------------------------
\section{Lense-Thirring precession}
\label{sec:lense}
%-----------------------------------------------------------------------------

In Sec.\,\ref{sec:test:GR}, we have only used the PK parameter $\dot{P}_{b}$ to
test gravity theories, but in our simulated results, the fractional error of
$\dot{\omega}$ could reach the highest precision compared with the other 4 PK
parameters.  So using $\dot{\omega}$ to test gravity is also powerful.  There is
another advantage of using $\dot{\omega}$ to test gravity, that the observed
$\dot{\omega}$ of an isolated binary pulsar system is relatively ``clean'',
because, in general, the non-intrinsic contribution to $\dot{\omega}$ can be
ignored.  So our simulated fractional error of $\dot{\omega}$ is probably more
faithful.  In this section we provide a possible application of the high
precision $\dot{\omega}$ measurement, just for an illustrative purpose.

In relativistic gravity theories, different from Newtonian gravity, a rotating
body in a binary system can have an effect on spin and orbital dynamics.  For
PSR-NS systems, two  components all have proper rotation, and in leading order
contributions we only consider the effect from the spin-orbital (SO) coupling
\citep{Wex:2014nva}.  The relativistic SO coupling in PSR-NS systems can lead
the orbital and the two spins to precess.  For the precession of spin, the main
effect is a secular changes in the orientation of spin and this effect is due to
the space-time curvature and is spin-independent.  We generally call this effect
geodetic precession and it has been detected in some binary pulsar systems, such
as PSR J0737$-$3039 \citep{Breton2008}.  On the other hand, the precession of
orbit contains two effects which needs to be considered.  The first effect is
due to the mass-mass interaction.  The second effect is due to the
Lense-Thirring effect related to the spins of two stars. 

\citet{1988NCimB.101..127D} decomposed the effect of precession of orbit in
terms of observable parameters of pulsars and calculated the contribution to periastron advance $\dot{\omega}_{p}$,
%--
\begin{align}\label{eq:omegalense}
    \dot{\omega}_{p}=\dot{\omega}_{\rm PN}+\dot{\omega}_{{\rm LT}_{p}}+\dot{\omega}_{{\rm LT}_{c}}\,,
\end{align}
%--
where $\dot{\omega}_{\rm PN}$ is caused by the mass-mass interaction which
results from the two bodies of binary system, $\dot{\omega}_{{\rm LT}_{p}}$
and $\dot{\omega}_{{\rm LT}_{c}}$ are due to the Lense-Thirring effect of the
pulsar and the companion respectively.  For the actual observed value of
$\dot{\omega}^{\rm obs}_{p}$, it should include the \emph{Kopeikin term} which
is caused by the proper motion of a binary system \citep{Kopeikin_1996}, but
in general, the contribution of \emph{Kopeikin term} is small and can be
ignored or subtracted \citep{Bagchi_2018,Hu:2020ubl}.  The expression of the
first term in Eq.\,(\ref{eq:omegalense}) is
\begin{align}
    \dot{\omega}_{\rm PN}=&3 T_{\odot}^{2 / 3}\left(\frac{P_{b}}{2 \pi}\right)^{-5 / 3} \frac{\left(m_{p}+m_{c}\right)^{2 / 3}}{1-e^{2}}\left(1+f_{\rm O}v^{2}\right)\,,\label{eq:omega2pn_all}\\
    f_{\rm O}=&\frac{1}{1-e^{2}}\left(\frac{3}{2}X_{p}^{2}+\frac{3}{2}X_{p}+\frac{27}{4}\right)+\left(\frac{5}{6}X_{p}^{2}-\frac{23}{6}X_{p}-\frac{1}{4}\right)\,,
\end{align}
where $X_{p} \equiv m_{p}/m$.  The first term on the right side of
Eq.~(\ref{eq:omega2pn_all}) is the 1 PN term of $\dot{\omega}_{\rm PN}$ and is
given by Eq.\,(\ref{eq:omegadot}); the second term is the 2 PN term of
$\dot{\omega}_{\rm PN}$.  We calculate the fractional value of the 2 PN term
for the 8-min PSR-NS system II and get $f_{O}v^{2}\approx2\times10^{-4}$, so
the contribution from $\dot{\omega}_{\rm 2PN}$ is four orders of magnitude
smaller than $\dot{\omega}_{\rm 1PN}$.  Our simulated fractional error of
$\dot{\omega}_{\rm 1PN}$ can reach $10^{-8}$ for an 8-min PSR-NS system with
$\sigma_{\rm TOA}=1\,\mu$s, so the contribution from $\dot{\omega}_{\rm 2PN}$
should be detectable for an 8-min binary.  
% But as we discussed in
% Sec.\,\ref{sec:sim}, the $\dot{\omega}_{\rm 2PN}$ is still a small value
% relative to $\dot{\omega}_{\rm 1PN}$, so it could not affect the magnitude
% estimate of the expected fractional uncertainty of $\dot{\omega}$. 

The expressions of the second and third terms in Eq.\,(\ref{eq:omegalense})
are \citep{Bagchi_2018,Hu:2020ubl},
\begin{align}
    &\dot{\omega}_{{\rm LT}_{p}}+\dot{\omega}_{ {\rm LT}_{c}}=-3 T_{\odot}^{2 / 3}\left(\frac{P_{b}}{2 \pi}\right)^{-5 / 3} \frac{\left(m_{p}+m_{c}\right)^{2 / 3}}{1-e^{2}}v\nonumber \\
    & \hspace{2.5cm} \times\left(g^{\rm s}_{p}\beta^{\rm s}_{p}+g^{\rm s}_{c}\beta_{c}^{\rm s}\right)
    \,,\\
    &\beta^{\rm s}_{p}=\frac{cI_{p}}{GM^{2}_{p}}\frac{2\pi}{P_{p}}\,,\label{eq:betasp}\\
    &g^{\rm s}_{p}=\frac{X_{p}\left(4 X_{p}+3 X_{c} \right)}{6 \left( 1-e^{2} \right)^{1/2}\sin^{2}{i}}\left[\left(3\sin^{2}{i}-1\right)\cos{\chi_{p}}+\cos{i}\cos{\lambda_{p}}\right]\,,\label{eq:gsp}
\end{align}
where $I_{p}$ is the moment of inertia of the pulsar and depends on the EOS of
NSs, and $P_{p}$ is the spin period of the pulsar.  $\chi_{p}$ is the angle
between $\mathbf{s}_{p}$ and $\mathbf{k}$, where $\mathbf{k}$ is the unit vector
along the orbital angular momentum and $\mathbf{s}_{p}$ is the unit spin vector
of the pulsar; $\lambda_{p}$ is the angle between $\mathbf{s}_{p}$ and
$\mathbf{h}_{p}$ where $\mathbf{h}_{p}$ is the unit vector along the
line-of-sight.  To get $\beta_{c}^{\rm s}$ and $g_{c}^{\rm s}$, we just need to
interchange the subscript ``$c$'' and ``$p$''.  Generally, the pulsar in a
PSR-NS system is a recycled pulsar and the companion is a normal pulsar, so the
rotation of companion is much slower than the pulsar, and we can ignore the
contribution of $\dot{\omega}_{{\rm LT}_{c}}$, see e.g. PSR~J0737$-$3039 \citep{Kramer:2009zza}.

To estimate the magnitude of $\dot{\omega}_{{\rm LT}_{p}}$ contributed by the
Lense-Thirring effect, we simplify the calculation a little bit.  If
$\mathbf{s}_{p}$ is parallel to $\mathbf{k}$, namely $\chi_{p}=0$ and
$\lambda_{p}=i$, Eq.\,(\ref{eq:gsp}) can be simplified to,
\begin{align}
    g_{p\,\parallel}^{\rm s}=\frac{\left(\frac{1}{3} X^{2}_{p}+ X_{p} \right)}{ \left( 1-e^{2} \right)^{1/2}}\,.
\end{align}
With these equations, we can estimate the contribution from $\dot{\omega}_{{\rm LT}_{p}}$ for periastron advance $\dot{\omega}_{p}$. For the 8-min PSR-NS system II,
\begin{align}
    \dot{\omega}_{\rm LT_{p}}=-0.14\,{\rm deg\,yr^{-1}}\left(\frac{I_{p}}{10^{45}\,{\rm g\,cm^{2}}}\right)\left(\frac{20\,{\rm ms}}{P_{p}}\right)\,.
\end{align}
So if $I_{p}=10^{45}\,{\rm g\,cm^{2}}$  and $P_{p}=20\,{\rm ms}$, we have
$\dot{\omega}_{\rm LT_{p}}=-0.14\,{\rm deg\,yr^{-1}}$.  For the 8-min PSR-NS
system II, $\left|\dot{\omega}_{\rm LT_{p}}/\dot{\omega}_{\rm
1PN}\right|\approx6.1 \times10^{-5}$. Remember that the fractional error of
$\dot{\omega}$ can reach $10^{-8}$ with $\sigma_{\rm TOA}=1\,{\rm \mu s}$ in
our simulation.  So the contribution of $\dot{\omega}_{ {\rm LT}_{p}}$ can be
significant compared with the error of $\dot{\omega}^{\rm obs}$, and it means
that we are very likely to detect the value of $\dot{\omega}_{ {\rm LT}_{p}}$
contributed from the Lense-Thirring effect.
Notice that the value of $\dot{\omega}_{ {\rm LT}_{p}}$ depends on the
moment of inertia $I_{p}$, as shown in Eq.\,(\ref{eq:betasp}).  When $m_{p}$
and spin are determined, $I_{p}$ only depends on the EOS.  So if we can
separate $\dot{\omega}_{ {\rm LT}_{p}}$, it would help us restrict the EOS.
This is another way to restrict the EOS in addition to using the maximum mass of NSs \citep{Lattimer_2005,Hu:2020ubl}. 

%---------------------------------------------------------------------
\section{Discussions}
\label{sec:disc}

In our work, we assume that one can use the LISA and SKA multimessenger
strategy \citep{Kyutoku:2018aai, Andrews:2019plw, Lau:2019wzw} to detect
short-orbital-period PSR-NS systems with $P_{b}\sim {\rm min}$ in the future.
The short-orbital-period PSR-NS systems can provide extreme gravity tests
which may show new information for gravity.  We use the plug-in FAKE of TEMPO2
to simulate the TOAs of short-orbital-period PSR-NS systems and fit TOAs to
get the fractional errors of 5 PK parameters.  Although we do not consider the
higher PN order contribution in our simulation, considering that $v \ll c$, the main contribution from the
lowest PN order can provide reliable errors of parameters which can
help us constrain gravity theories.  As our results show, compared with the
previous bounds from binary pulsar systems, short-orbital-period systems
can significantly improve the constraints on parameters of some specific
gravity theories.  The constraints of parameters we predicted are looser than
the bound from the Solar System, but {highly} relativistic PSR-NS systems  give
constraints from strong field which is a complementary gravity regime.  Our
simulation is simplified and we expect to provide a more complete simulation
in the future, such as to study a more complete timing model.

For our simulation, we choose two different groups of masses (see
Table~\ref{tab:PSR-NSsparameters}). We found that PSR-NS I system
($m_{p}=1.3\,M_{\odot}$, $m_{p}=1.7\,M_{\odot}$) can provide a slightly tighter
bound than PSR-NS II system ($m_{p}=1.35\,M_{\odot}$, $m_{p}=1.44\,M_{\odot}$)
with a same $\sigma_{\rm TOA}$.  It indicates that PSR-NS systems with larger
mass differences are more suitable for limiting these theories of gravity that
we discussed, in particular for the dipole radiation in the DEF gravity  where
the asymmetry of masses will play an important role { \citep[see e.g. ][in
the context of GWs]{Zhao:2021bjw}}.

We simulate 5 PK parameters, among them, $r$ and $s$ are parameters related to
the Shapiro delay.  The precisions of $r$ and $s$ depend on the orbital
inclination $i$ which we have set $i=60^{\circ}$ in our simulation.  If the PSR-NS
system is nearly edge-on ($i \sim 90^{\circ}$), the effect of Shapiro delay can be more
significant, such as in PSR J0737$-$3039 \citep{Kramer:2006nb}, and it can
make $r$ and $s$ more precise in actual measurement.  The higher precisions of
$r$ and $s$ can help us measure the masses of PSR-NS systems more precisely and
test GR more stringently.

In the future,  TianQin \citep{Luo:2015ght} and Taiji~\citep{Luo:2020tmp} could
also join the multimessenger observation strategies.  The sensitive frequency of
Taiji is slightly lower than LISA, which allows the joint observation of Taiji
and SKA to detect more PSR-NS systems with a bit larger ${P}_{b}$.  In addition,
the mission times of TianQin or Taiji might be different from LISA, so in that
case we can have a longer total time to detect short-orbital-period PSR-NS
systems by different combinations of joint observation strategies. This may
increase the detection numbers.

In the astrophysical side, if we can detect short-orbital-period PSR-NS systems
in the future, it can fill in the gap in current observations and make us better
model the DNS population.  Detecting such systems can help us learn the
formation of DNSs better, and possibly verify new formation channels of DNSs,
such as a fast-merging channel \citep{Andrews:2019plw}.

Furthermore, if LISA and SKA can detect a PSR-NS system, as we discussed
earlier, LISA can help SKA to locate and detect PSR-NS systems. Meanwhile, SKA
can provide a more accurate measurement of parameters, which in turn, could help
us analyse the GW waveforms of  PSR-NS systems in the LISA data. We leave these
studies to future publication.

%---------------------------------------------------------------------
%---------------------------------------------------------------------
\acknowledgments

{We thank the anonymous referee for constructive comments that improved the
work.} This work was supported by the National SKA Program of China
(2020SKA0120300), the National Natural Science Foundation of China (11975027,
11991053, 11721303), the Young Elite Scientists Sponsorship Program by the China
Association for Science and Technology (2018QNRC001), and the Max Planck Partner
Group Program funded by the Max Planck Society.  It was partially supported by
the Strategic Priority Research Program of the Chinese Academy of Sciences
(XDB23010200), and the High-Performance Computing Platform of Peking University.
%---------------------------------------------------------------------

\vspace{5mm}
\facilities{LISA, SKA}

\software{TEMPO2 \citep{Hobbs_2006,Edwards_2006}}

% \clearpage

%---------------------------------------------------------------------
\bibliography{refs}
%--------------------------------------------------------------------

\end{document}